\documentclass[12pt]{article}
\usepackage{amssymb,amsmath,latexsym}
\usepackage{graphicx}
\usepackage[final]{pdfpages}
\usepackage{verbatim}
\usepackage{color}
\usepackage{amsfonts,amsthm,hyperref,mathrsfs,ulem,tikz}

\usetikzlibrary{arrows,patterns}

\title{
Pseudo-gaps for random hopping models}

\author{Florian Dorsch, Hermann Schulz-Baldes
\\
\\
{\small Department Mathematik, Friedrich-Alexander-Universit\"at Erlangen-N\"urnberg, Germany}
}

\date{ }

\newtheorem{theo}{Theorem}
\newtheorem{defini}[theo]{Definition}
\newtheorem{proposi}[theo]{Proposition}
\newtheorem{lemma}[theo]{Lemma}

\newcommand{\ZZ}{{\mathbb Z}}

\newcommand{\RR}{{\mathbb R}}

\newcommand{\CM}{{\mathbb C}}
\newcommand{\NM}{{\mathbb N}}
\newcommand{\RM}{{\mathbb R}}

\newcommand{\ZM}{{\mathbb Z}}

\newcommand{\PP}{{\bf P}}

\newcommand{\EE}{{\bf E}}

\newcommand{\Oo}{{\cal O}}

\newcommand{\Nn}{{\cal N}}

\def\essinf{\mathop{\rm ess\,inf}}
\def\esssup{\mathop{\rm ess\,sup}}

\newcommand{\one}{{\bf 1}}

\newcommand{\ev}{{\mbox{\rm\tiny ev}}}
\newcommand{\odd}{{\mbox{\rm\tiny od}}}
\newcommand{\kConst}{k}
\newcommand{\KConst}{K}
\newcommand{\bareT}{\widehat{T}}
\newcommand{\baret}{\hat{t}}
\newcommand{\barev}{\hat{v}}

\newcommand{\bsm}{\left(\begin{smallmatrix}} 
\newcommand{\esm}{\end{smallmatrix}\right)}  
\definecolor{GR}{rgb}{.35,.7,.35}

\begin{document}

\maketitle

\begin{abstract}
For one-dimensional random Schr\"odinger operators, the integrated density of states is known to be given in terms of the (averaged) rotation number of the Pr\"ufer phase dynamics. This paper develops a controlled perturbation theory for the rotation number around an energy at which all the transfer matrices commute and are hyperbolic. Such a hyperbolic critical energy appears in random hopping models. The main result is a H\"older continuity of the rotation number at the critical energy that implies the existence of a pseudo-gap. The proof uses renewal theory. The result is illustrated by numerics.


\end{abstract}


\vspace{.5cm}


\section{Intuition and main result}
\label{sec-intro}

The main result of this note and the intuition behind it can directly be explained by looking at a concrete situation. A more general theoretical approach is deferred to the subsequent sections. A random hopping model is a discrete random Schr\"odinger operator on the Hilbert space $\ell^2(\ZM)$ of the form
\begin{equation}
\label{eq-IntroHam}
(H\psi)(n)
\;=\;
-\,t(n+1) \psi(n+1)\,-\,t(n)\psi(n-1)
\;,
\qquad
\psi\in\ell^2(\ZZ)
\;,
\end{equation}
where $(t(n))_{n\in\ZM}$ is a sequence of independent positive random variables. The model has a bipartite chiral symmetry, namely $JHJ=-H$ for the operator $J|n\rangle=(-1)^{n}|n\rangle$ which is a symmetry in the sense that $J=J^*$ and $J^2=\one$. This implies, in particular, that the spectrum and density of states is symmetric around the energy $0$. For special choices of the distribution, the model is the random Hu-Seeger-Schriefer model \cite{MSHP} as well as a model that maps to certain quantum spin chains \cite{CS}. A standard way to rewrite the Schr\"odinger equation $H\psi=E\psi$ for a real energy $E\in\RM$ is to use the transfer matrices
$$
\bareT^E_n
\;=\;
\begin{pmatrix}
-\,E\,\tfrac{1}{t(n)} & -t(n) \\ \tfrac{1}{t(n)} & 0
\end{pmatrix}
\;.
$$
For $E$ sufficiently small and for $t(n)$ bounded away from $0$, these matrices are elliptic, namely conjugate to a rotation matrix. Our focus will be on a situation where the $t(n)$ are independent random variables that have the same distribution for even and odd $n$, respectively. Then it is natural to consider the transfer matrices over dimers, that is, the product of two adjacent matrices:
$$
T^E_n
\;=\;
\bareT^E_{2n+1}\bareT^E_{2n}
\;=\;
\begin{pmatrix}
E^2\tfrac{1}{t(2n+1)t(2n)} - \tfrac{t(2n+1)}{t(2n)} & E\,\tfrac{t(2n)}{t(2n+1)} \\ -\, E \,\tfrac{1}{t(2n+1)t(2n)} & -\tfrac{t(2n)}{t(2n+1)}
\end{pmatrix}
\;.
$$
At $E=0$, the matrices $T^0_n$ are all diagonal and thus commute, and furthermore, unless $t(2n)=t(2n+1)$, the matrices all have a trace of modulus larger than $2$ and are thus hyperbolic with two eigenvalues off the unit circle. More generally (see below), an energy with commuting hyperbolic (polymer) transfer matrices is called a hyperbolic critical energy. One can expand $T^E_n$ in small energies $E$ as follows
\begin{align}
\label{eq-GenDyn}
T^E_n
& 
\;=\;
-\,
\left[
\one
\;+\;
E 
\begin{pmatrix}
0 & -1 \\ \tfrac{1}{t(2n+1)^2} & 0
\end{pmatrix}
\;-\;
E^2
\begin{pmatrix}
\tfrac{1}{t(2n+1)^2} & 0 \\ 0 & 0
\end{pmatrix}
\right]
\begin{pmatrix}
\tfrac{t(2n+1)}{t(2n)} & 0 \\ 0 & \tfrac{t(2n)}{t(2n+1)}
\end{pmatrix}
\;,
\end{align}
namely up to errors $T^E_n$ is the product of a random diagonal hyperbolic matrix and a matrix close to the identity which is, up to fluctuations, a rotation of order $E$. Next let us recall the associated dynamics on the Pr\"ufer phases $\theta$ specifying a unit vector and hence a direction in $\RM^2$ via the notation 
$$
e_\theta\;=\;\begin{pmatrix} \cos(\theta) \\ \sin(\theta)\end{pmatrix}
\;.
$$ 
The action on these phases is defined iteratively by 
$$
R^E_n\, e_{\theta^E(n)}\;=\; 
T^E_n\,e_{\theta^E(n-1)}
\;,
$$ 
where $R^E_n$ is some normalization constant and $\theta^E(0)$ some initial condition. Under the stereographic projection, this becomes the M\"obius action of the cotangent of the Pr\"ufer phases:
\begin{equation}
\label{eq-PrueferDyn}
\cot(\theta^E(n))
\;=\;
\frac{t(2n+1)^2}{t(2n)^2}
\;
\frac{\cot(\theta^E(n-1))(1-E^2\,\tfrac{1}{t(2n+1)^2}) \, -\,E \,\tfrac{t(2n)^2}{t(2n+1)^2} }{1\,+\,E\,\tfrac{1}{t(2n)^2} \cot(\theta^E(n-1)) }
\;.
\end{equation}
As the cotangent is $\pi$-periodic, this equation can be read as a dynamics on $(-\frac{\pi}{2},\frac{\pi}{2}]$ which reflects that the direction of $e_\theta$ is fixed by the value of $\theta$ in the projective space isomorphic to $(-\frac{\pi}{2},\frac{\pi}{2}]$ (later on the dynamics will be lifted to an action on $\RM$).  For $E=0$, the dynamics simply reduces to $\cot(\theta^0(n))=\kappa(n)^2\cot(\theta^0(n-1))$ where $\kappa(n)=\frac{t(2n+1)}{t(2n)}$. On the unit circle this becomes
$$
e^{2\imath\theta^0(n)}
\;=\;
\frac{(\kappa(n)+\tfrac{1}{\kappa(n)})\,e^{2\imath\theta^0(n-1)}+(\kappa(n)-\tfrac{1}{\kappa(n)})}{(\kappa(n)-\tfrac{1}{\kappa(n)})\,e^{2\imath\theta^0(n-1)}+(\kappa(n)+\tfrac{1}{\kappa(n)})}
\;.
$$
Independent of $\kappa(n)$, this dynamics has two fixed points at $\theta=0$ and $\theta=\frac{\pi}{2}$. For $\kappa(n)>1$, $\theta=0$ is attractive and $\theta=\frac{\pi}{2}$ is repulsive, and visa versa for $\kappa(n)<1$. Now let us consider a situation where the $\kappa(n)$ are i.i.d. with random positive values that can be either larger or smaller than $1$. In the average, this dynamics may lead to a drift to $e_0$ or $e_{\frac{\pi}{2}}$, pending on the distribution. This drift is actually dictated by the Lyapunov exponent at $E=0$:
\begin{figure}
\centering
\includegraphics[width=5.7cm]{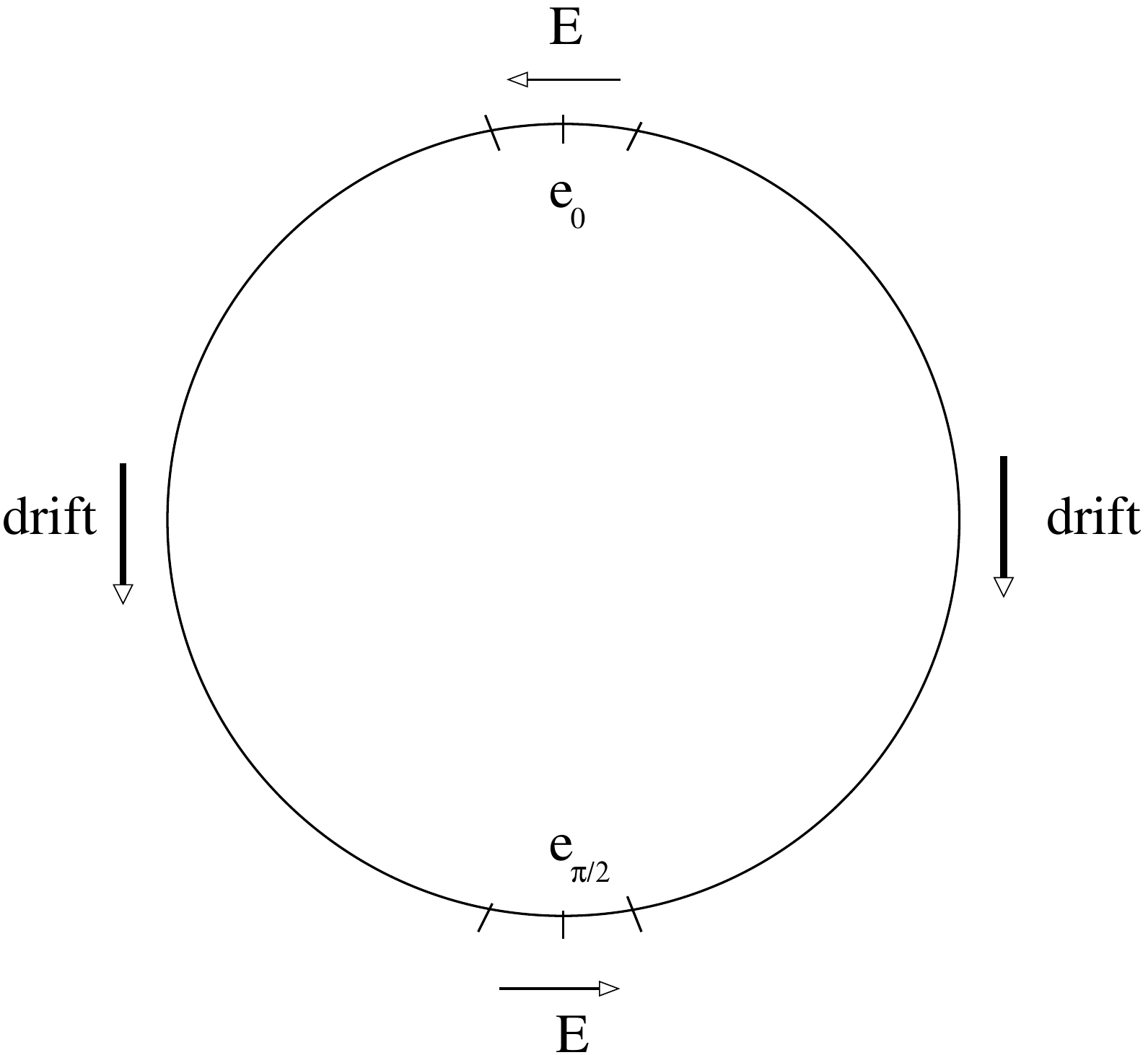} 
\hspace{.8cm}
\includegraphics[width=8cm]{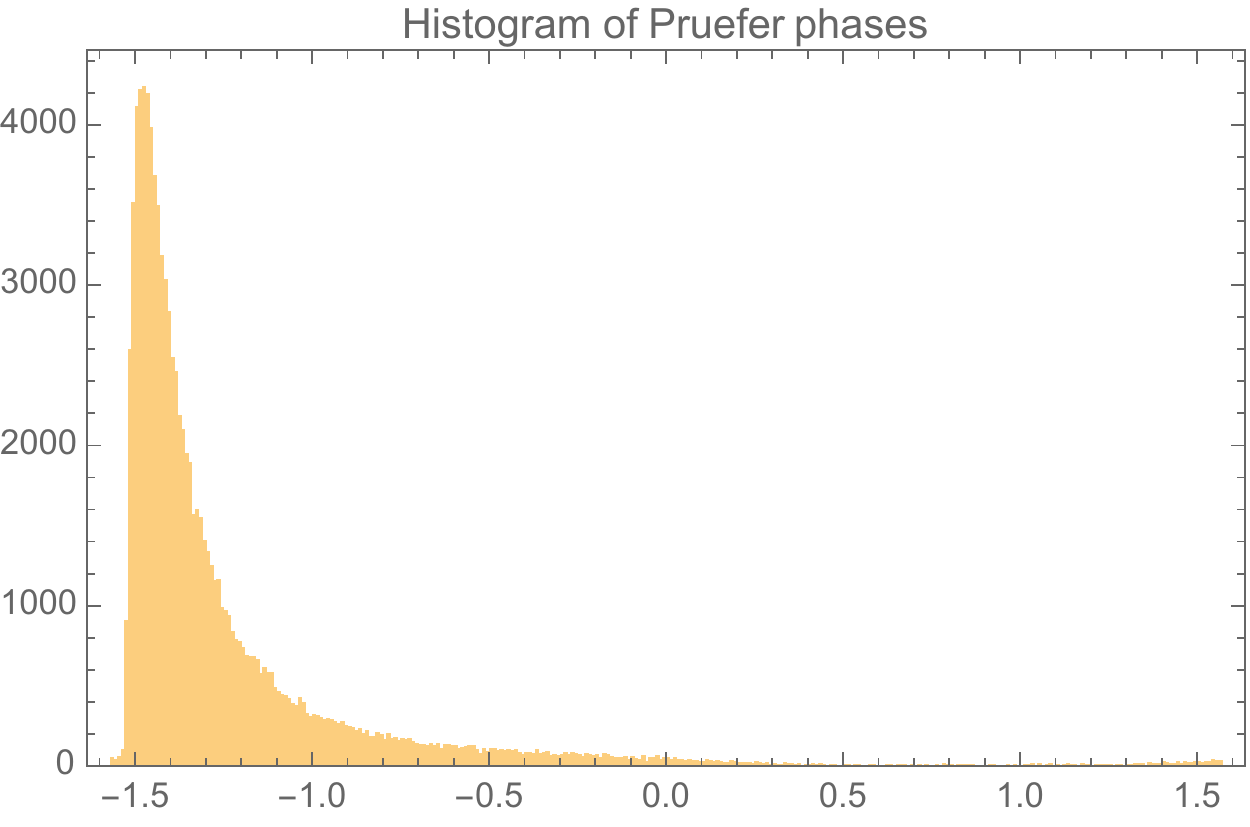} 
\caption{Schematic representation of the random dynamics as described in the text. The histogram shows the distribution of $10^5$ Pr\"ufer phases generated by \eqref{eq-PrueferDyn} with a distributions of the hopping terms given by \eqref{eq-HoppingDistr} where $x$ is uniformly distributed in $[-1,1]$. The parameters are $c_\ev=1.2$, $\lambda_\ev=0.4$, $c_\odd=1$ and $\lambda_\odd=0$ so that $\gamma^0<0$, and the energy $E=0.02$. The weight on the right half (in $[0,\frac{\pi}{2}]$) results from $101$ rotations during the $10^5$ random dynamical steps.
}
\label{fig-DynSymbolic}
\end{figure}
$$
\gamma^0
\;=\;
\lim_{N\to\infty}\;\frac{1}{N}\;\bigg\langle\log\bigg(\prod\limits_{n=1}^{N}\kappa(n)\bigg)\bigg\rangle
\;=\;
\big\langle\log\big(\kappa(n)\big)\big\rangle
\;.
$$
It dictates the growth of the upper component of \eqref{eq-GenDyn} at $E=0$. The lower component has a Lyapunov exponent $-\gamma^0$. If now $\gamma^0>0$, then there is a drift to $e_0$, while for $\gamma^0<0$ the drift is to $e_{\frac{\pi}{2}}$. This latter is the case in Fig.~\ref{fig-DynSymbolic} and we restrict to this case for the moment. The case $\gamma^0=0$ is not considered in this work. 

\vspace{.2cm}

Now let us consider the energy dependent part in \eqref{eq-GenDyn}. Of importance is that the two signs in the linear term in $E$ are independent of the distribution of the $t(n)$. Hence the second factor is, up to corrections, a rotation by a random phase of order $E$ in the positive orientation for $E>0$. While almost everywhere on the circle this rotation is very small compared to the hyperbolic dynamics generated by the hyperbolic factor $\kappa(n)^2$, it is dominant close to the two fixed points $e_0$ and $e_{\frac{\pi}{2}}$. This is also included in Fig.~\ref{fig-DynSymbolic}. Finally we can sketch intuitively the behavior of the random dynamics. Suppose one starts in a neighborhood of $e_0$, either to the left or the right. In such a neighborhood, the hyperbolic dynamics is ineffective (recall that $e_0$ is a fixed point for all $\kappa(n)$), however, there is a counter-clockwise rotation by random phases. Eventually, the dynamics will leave the neighborhood and get into a region where the hyperbolic dynamics is effective. Due to the drift (see again Fig.~\ref{fig-DynSymbolic}) the Pr\"ufer phase typically reaches a neighborhood of $e_{\frac{\pi}{2}}$ after a finite number of steps. Again this neighborhood is crossed counter-clockwise due to the random rotations. Finally, the dynamics reaches the r.h.s. of the circle (projective space). Here it faces a drift which presses it back towards $e_{\frac{\pi}{2}}$ which, however, it cannot cross backwards due to the counter-clockwise rotations at $e_{\frac{\pi}{2}}$. Hence the Pr\"ufer phase is for many iterations bound to stay close to the right of $e_{\frac{\pi}{2}}$, see the histogram in Fig.~\ref{fig-DynSymbolic}. The only way to reach $e_0$ is via rare sequences of values $\kappa(n)>1$. To analyze the corresponding large deviations is a crucial element of understanding the random dynamics. Clearly, if the sign of $E$ and $\gamma^0$ change, the schematic representation changes (orientation and the respective roles of $e_0$ and $e_{\frac{\pi}{2}}$), but the heuristics and therefore also the arguments below are the same. Throughout all arguments we focus only on the case $\gamma^0<0$ and $E>0$.

\vspace{.2cm}

Clearly, from the dynamical point of view it is of interest to study the random times needed to make a loop around projective space. Each time the dynamics passes by $e_0$ (or alternatively $e_{\frac{\pi}{2}}$) the process starts anew. Therefore summing all random loop times is precisely what is called a renewal process. The elementary renewal theorem (see below) links the average time to make a loop to the inverse of the expected value of the time needed for one loop. The average time to make a loop is also called the rotation number and it is well-known that it is equal to the integrated density of states  (IDS) of the random Schr\"odinger operator which is the non-decreasing function $E\in\RM\mapsto \Nn(E)$  defined by
$$
\Nn(E)
\;=\;
\lim_{N\to\infty}\frac{1}{N}\;\#\{\mbox{\rm eigenvalues of }H_N\,\leq\,E\}
\;,
$$
where $H_N$ is the restriction of $H$ to $\ell^2(\{1,\ldots,N\})$.  The limit is known to exist almost surely. The IDS is one of the most basic quantities describing a random Schr\"odinger operator and its continuity properties are of great importance. Connecting it to the rotation number of the Pr\"ufer phases requires some care and this is done in Section~\ref{sec-RotNumers}. Then using the detailed information on the Pr\"ufer phase dynamics and its dependence on parameters like the energy $E$ and the distribution of the $t(n)$ allows to prove a new result on the IDS at a hyperbolic critical energy, such as $E=0$ in the random hopping model described above. The following theorem shows that there is an exponent $\nu$ depending on the distribution which provides a H\"older estimate on the IDS at the critical energy. This exponent $\nu$ can easily be made very large and then the result implies that there is a characteristic pseudo-gap in the IDS, namely the DOS vanishes at the critical energy with a large H\"older exponent. Figure 2 provides a striking numerical example for this.

\begin{theo}\label{main_result}
Suppose that the $t(n)$ are compactly supported in $(0,\infty)$ and such that $\gamma^0<0$. Moreover, assume that the $T_n^E$ are independent and identically distributed and such that the probability of having $\kappa(n)>1$ is positive. Then there exists a unique positive number $\nu$~obeying
\begin{equation}
\label{eq-KappaCrit}
\langle \kappa(n)^{\nu}\rangle
\;=\;1\,.
\end{equation}
For all $\delta>0$ there exists $C_\delta<\infty$ such that the integrated density of states satisfies
\begin{align}\label{main_result_bounded_statement}
\left|\mathcal{N}(E)-\mathcal{N}(0)\right|\;\leq\;C_\delta \;|E|^{\nu-\delta}
\;.
\end{align}
\end{theo}

Let us stress that if all $t(n)$ are i.i.d. one clearly has $\gamma^0=0$ so that the hypothesis of the theorem is not satisfied. On the other hand, having different distributions for even and odd sites generically leads to $\gamma^0\not= 0$ so that one can generate numerous examples in this manner. If this is guaranteed, the convexity of $\xi\mapsto \langle \kappa(n)^{\xi}\rangle$  implies the existence and uniqueness of $\nu$ which is positive for $\gamma^0<0$ and negative for $\gamma^0>0$ (note that $\gamma^0$ is the derivative of $\xi\mapsto \langle \kappa(n)^{\xi}\rangle$ at $\xi=0$). In particular situations it is possible to show that the bound \eqref{main_result_bounded_statement} is optimal, but we have not analyzed this in detail.

\vspace{.2cm}

Pseudo-gaps as \eqref{eq-KappaCrit} with $\nu>1$ appear in numerous models of solid state physics. They can result from interactions in high-$T_c$ superconductors \cite{LQS} or in non-interacting models of semimetals such as graphene \cite{Neto}. Furthermore, also certain quasi-one-dimensional Bogoliubov-de Gennes Hamiltonians have pseudo-gaps \cite{TBFM}. In these two latter cases, symmetries play a crucial role. Also in the model leading to \eqref{eq-GenDyn} there is a chiral symmetry (related to the bipartite structure). Nevertheless,  to our best knowledge, there are no earlier works on pseudo-gaps in strictly one-dimensional random models. Moreover, the remainder of the paper shows how to construct such models with a pseudo-gap. 

\vspace{.2cm}

Let us also point out that exponents $\nu$ defined in a similar manner to \eqref{eq-KappaCrit} played  a role in \cite{DH,GGG,HA}. These papers looked at the Lyapunov exponent near a critical value (corresponding to a critical energy in our terminology described below) and exhibited singular behavior of the Lyapunov exponent in its vicinity, namely a deviation from the standard quadratic vanishing of the Lyapunov exponent. A key role in the analysis in \cite{DH}, and its rigorous version \cite{GGG}, is a perturbative control of the invariant Furstenberg measures. Given the tight connection between the IDS and Lyapunov exponent via the Thouless formula, it is hence not surprising that also the IDS can have a singular behavior as in \eqref{main_result_bounded_statement}. This has not been worked out elsewhere though, again as far as we know. In fact, a difficulty is linked to the non-local nature of the Thouless formula: an information on the scaling of either IDS or Lyapunov exponent at one point (the critical value) does not allow to deduce information about the other. For example, to establish H\"older regularity of the Lyapunov exponent (as in \cite{DK}) requires H\"older regularity of the IDS in a neighborhood of the critical energy, and not just the pointwise information \eqref{main_result_bounded_statement}. In this paper, we do not argue based on the Thouless formula, but rather use oscillation theory to access the IDS directly.

\vspace{.2cm}

The remainder of the paper is organized as follows. The short next section presents and discusses some numerical results that illustrate Theorem~\ref{main_result}. Section~\ref{sec-RotNumers} presents the general framework of random polymer models (essentially based on \cite{JSS}) and then defines the notion of hyperbolic critical energy (different from the type of critical energies analyzed in \cite{JSS}). This singles out the main structural features of a random Jacobi matrix that lead to a Pr\"ufer phase dynamics as qualitatively described in Fig.~1 and thus also a behavior of the IDS as in Theorem~\ref{main_result}. Section~\ref{sec-MainTechnics} then contains the core of the mathematical analysis. In particular, deterministic geometric arguments allow to connect the rotation number to renewal theory in Subsections~\ref{sec-Deterministic} and \ref{sec-renewal}, and in Subsection~\ref{sec-LD} the interarrival time is then estimated by a large deviation argument. Finally,  Subsection~\ref{sec-bounded} states and proves Theorem~\ref{upper_bound_rotation_number_bounded}, the most general result on pseudo-gaps in the framework of random polymer models. It incorporates Theorem~\ref{main_result}. The final Subsection~\ref{sec-unbounded} comments on how to extend the techniques to deal with random variables with unbounded support.

\section{Examples and numerical illustration}

This section illustrates Theorem~\ref{main_result} with several examples.  As already explained above, an interesting situation only appears if the even and odd sites of the random hopping model have different distributions. We suppose them to be of the following type
\begin{equation}
\label{eq-HoppingDistr}
t(2n)
\;\overset{{\rm d}}{=}\;
c_\ev\,+\,\lambda_\ev\, x\;,
\qquad
t(2n+1)
\;\overset{{\rm d}}{=}\;
c_\odd\,+\,\lambda_\odd \,x\;,
\end{equation}
where $\lambda_\ev<c_\ev$ and $\lambda_\odd<c_\odd$ are all positive parameters and $x$ is a random variable with values in $[-1,1]$. Hence all even sites have the same distribution, and so do all odd sites. Furthermore, all sites are supposed to be independent. Clearly one of the $4$ parameters (say the average of $c_\ev$ and $c_\odd$) is merely an energy scale and thus irrelevant.  To produce a non-trivial situation in the spirit of Theorem~\ref{main_result}, it is furthermore sufficient to just have randomness of say the even sites, which is achieved by choosing $\lambda_\odd=0$. This particular situation is of interest for the study of certain random quantum spin chains \cite{CS}. 

\vspace{.2cm}

\begin{figure}
\centering
\includegraphics[width=7cm]{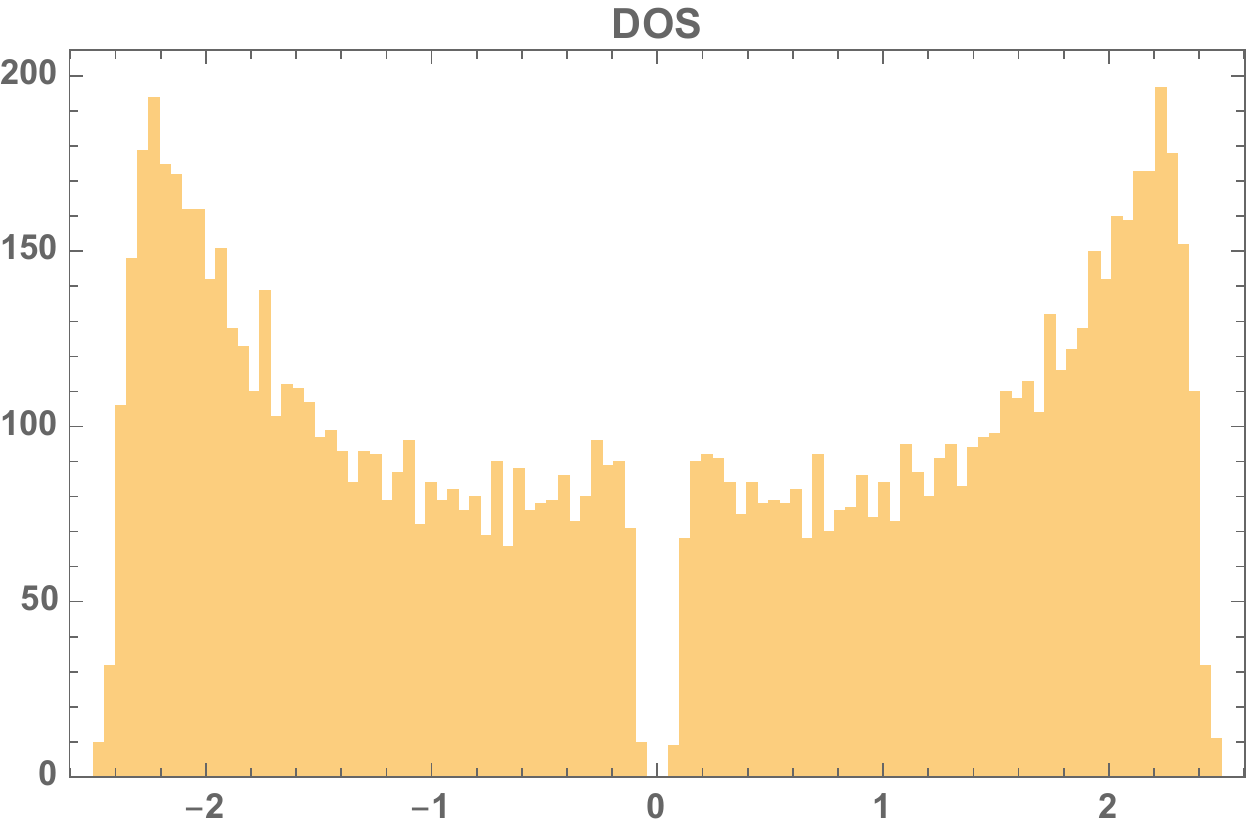}
\hspace{.3cm}
\includegraphics[width=8.5cm]{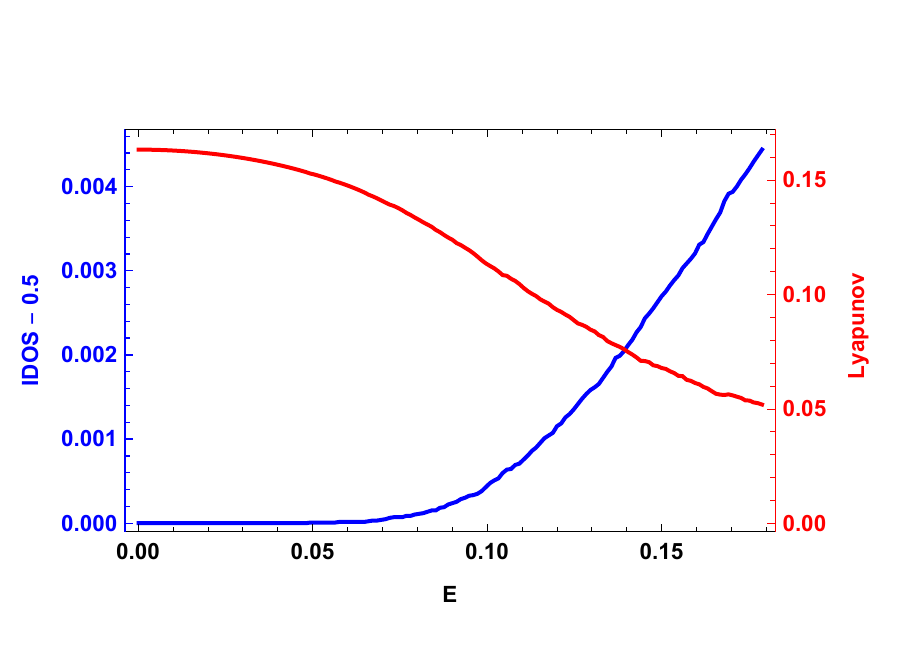}
\caption{The left figure shows a histogram of all eigenvalues of a random realization of the Hamiltonian of length $5000$ with same parameters as in Fig.~1. Up to normalization this is the density of states. The exponent in \eqref{eq-KappaCrit} is $\nu\approx 9.71$ which leads to the pseudo-gap. The right figure shows the integrated density of states close to $E=0$ as calculated numerically from the rotation number via \eqref{eq-IDSRotnumber}. }
\label{fig-SpecH2}
\end{figure}

The above model is also the Hu-Seeger-Schrieffer model if the odd sites are interpreted as random masses and the even ones as random hoppings between dimers. This model has a rich phase diagram \cite{MSHP} in the various parameters with quantum phase transitions at values of vanishing Lyapunov exponent $\gamma^0$ at zero energy. This is precisely the situation {\it not} analyzed in this paper. 

\vspace{.2cm}

As to the distribution of the random variable $x$, we consider two cases. In the first example, it is the uniform distribution on $[-1,1]$. In this case, one can evaluate explicitly
$$
\langle \kappa(n)^{\xi}\rangle
\;=\;
\frac{1-\xi}{1+\xi}\;
\frac{(c_\ev+\lambda_\ev)^{1+\xi}-(c_\ev-\lambda_\ev)^{1+\xi}}{\lambda_\ev}\;
\frac{\lambda_\odd}{
(c_\odd+\lambda_\odd)^{1-\xi}-(c_\odd-\lambda_\odd)^{1-\xi}}
\;.
$$
Note that one can take the limits $\lambda_\ev\to 0$ and $\lambda_\odd\to 0$. The solution to \eqref{eq-KappaCrit} can now readily be computed numerically. Furthermore, the zero energy Lyapunov exponent can be calculated (see \cite{MSHP}):
$$
\gamma^0
\;=\;
\langle \log(\kappa(n))\rangle
\;=\;
\frac{1}{2}\;\log
\left(
\frac{(c_\odd+\lambda_\odd)^{\frac{c_\odd}{\lambda_\odd}+1}}{
(c_\odd-\lambda_\odd)^{\frac{c_\odd}{\lambda_\odd}-1} }
\;
\frac{(c_\ev-\lambda_\ev)^{\frac{c_\ev}{\lambda_\ev}-1}}{(c_\ev+\lambda_\ev)^{\frac{c_\ev}{\lambda_\ev}+1}}
\right)
\;.
$$
A remarkable treat of these formulas is that the root $\nu$ of \eqref{eq-KappaCrit} strongly depends on the parameters of the model. 
A numerical evaluation of the global DOS and the IDOS and Lyapunov exponent value close to $E=0$ is provided in Figure 2.

\vspace{.2cm}

\begin{figure}
\centering
\includegraphics[width=5cm]{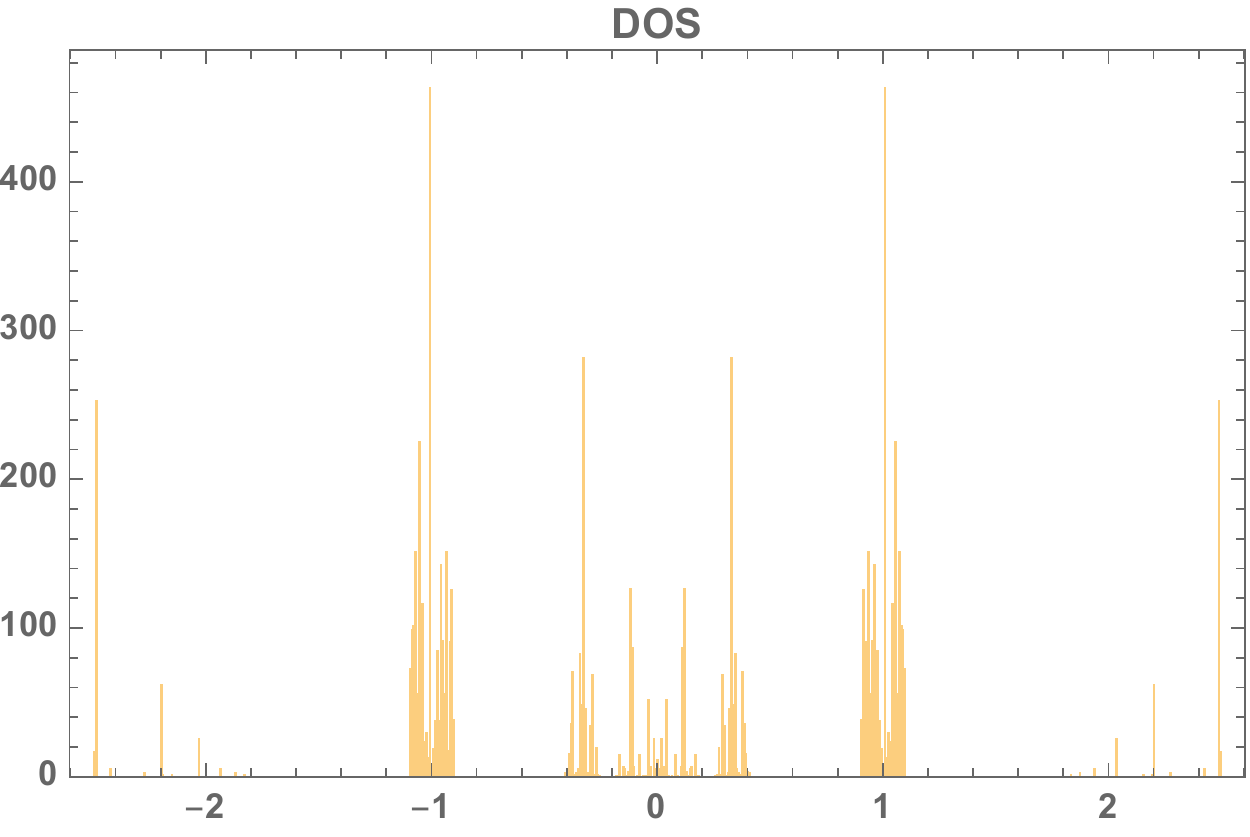}
\hspace{.3cm}
\includegraphics[width=5cm]{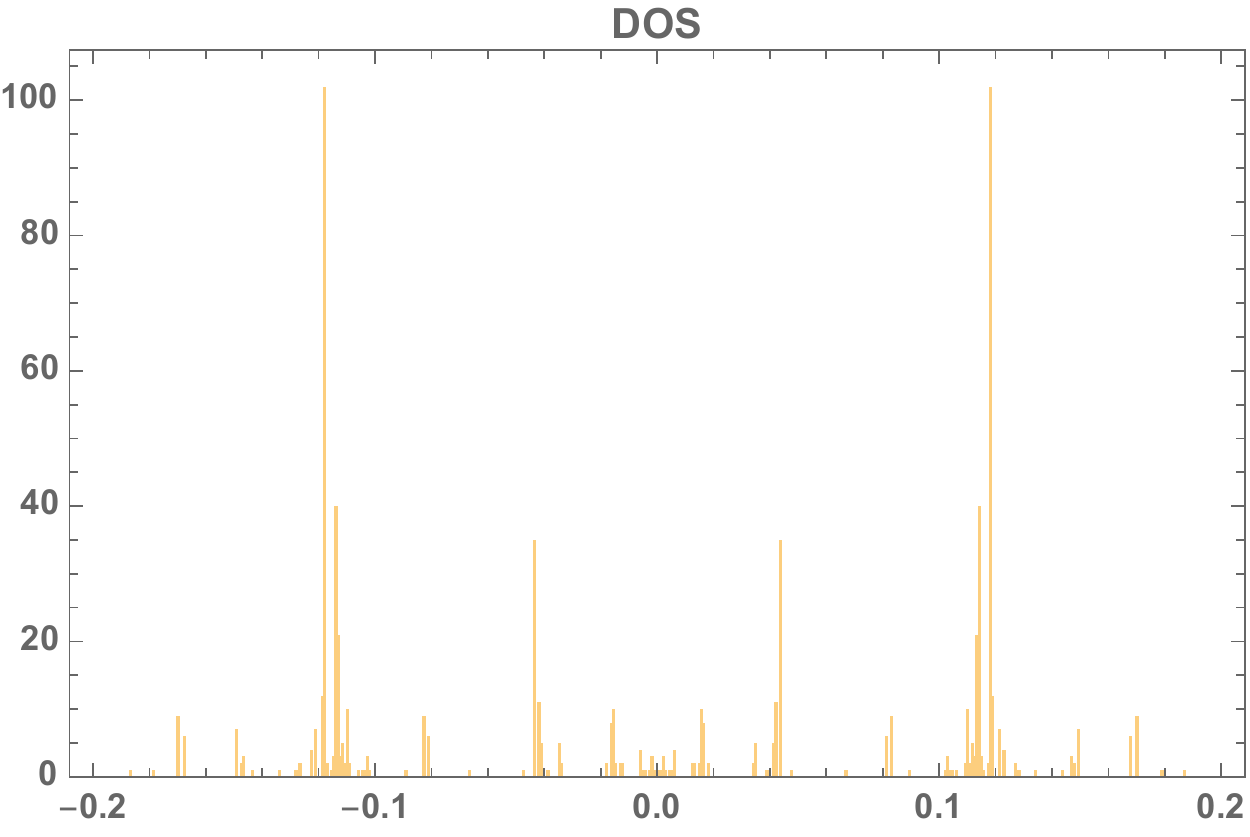}
\hspace{.3cm}
\includegraphics[width=5cm]{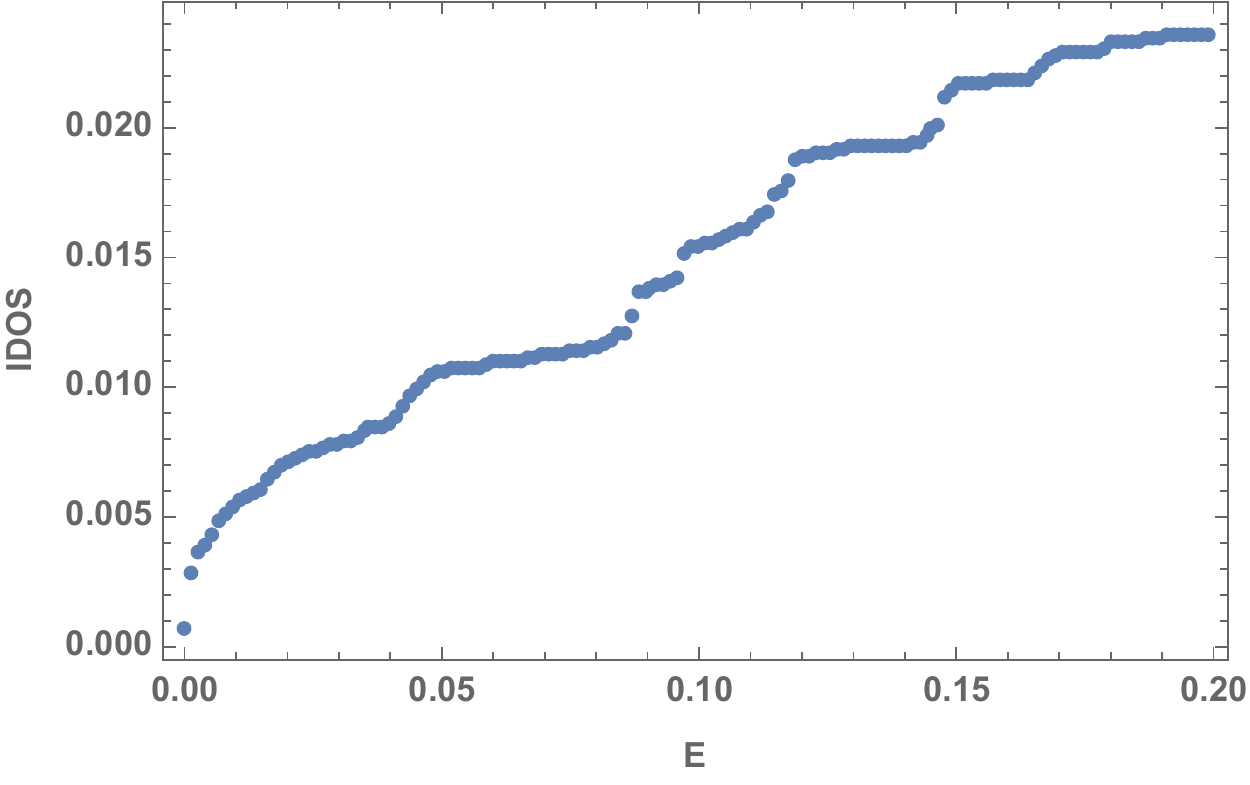}
\caption{For these graphs, the parameters are $c_\ev=1.4$, $\lambda_\ev=1.3$, $c_\odd=1$ and $\lambda_\odd=0$, and the even hopping terms were drawn with the Bernoulli distribution with  $p=\frac{2}{3}$ for the value~$1$. For these values, $\nu\approx 0.09$. The first two figures show the histogram of the eigenvalues, the second one being simply a zoom in the first one, and the third graph shows the IDS close to $E=0$  as calculated via the rotation number \eqref{eq-IDSRotnumber}. }
\label{fig-SpecH3}
\end{figure}

The second example considered here is that $x$ has the Bernoulli distribution $(1-p)\delta_{-1}+p\delta_1$ with parameter $p\in[0,1]$. Again it is possible to write out explicit formulas for $\langle \kappa(n)^{\xi}\rangle$ and $\gamma^0$, {\it e.g.},
$$
\langle \kappa(n)^{\xi}\rangle
\;=\;
\left(p(c_\odd+\lambda_\odd)^\xi\,+\,(1-p)(c_\odd-\lambda_\odd)^\xi\right)
\left(p(c_\ev+\lambda_\ev)^{-\xi}\,+\,(1-p)(c_\ev-\lambda_\ev)^{-\xi}\right)
\;.
$$
Bernoulli variables are known to easily lead to singular spectra. Indeed, this appears to be the case for the parameters chosen in Fig.~\ref{fig-SpecH3}. Furthermore, the spectrum surprisingly has some sort of self-similar structure. In this situation $\nu\approx 0.09$ is much smaller than $1$, leading to clustering of eigenvalues close to $E=0$.

\section{Rotation numbers at hyperbolic critical energies}
\label{sec-RotNumers}

\subsection{Polymer models and hyperbolic critical energies}

Let $\Sigma$ be a subset of $\bigcup\limits_{K=1}^L\{K\}\times\mathbb{R}_+^K\times\mathbb{R}^K$ where $L$ is a fixed maximal length. Any point $\sigma\in\Sigma$ is of the form $\sigma=(K,\baret_{\sigma}(0),\ldots,\baret_{\sigma}(K-1),\barev_{\sigma}(0),\ldots,\barev_{\sigma}(K-1))$ and fixes what we call a polymer (as in \cite{JSS}) of length $L_\sigma=K$ with hopping terms $\baret_\sigma=(\baret_{\sigma}(0),\ldots,\baret_{\sigma}(K-1))$ and potentials $\barev_\sigma=(\barev_{\sigma}(0),\ldots,\barev_{\sigma}(K-1))$.
Then let us consider the Tychonov space $\Omega_0=\Sigma^\ZM$. If ${\bf p}$ is a probability on $\Sigma$, then ${\bf P}_0={\bf p}^{\times \ZM}$ is a probability on $\Omega_0$ which is invariant and ergodic under the left shift $\tau_0:\Omega_0\rightarrow\Omega_0$ given by $(\sigma_m)_{m\in\mathbb{Z}}\rightarrow(\sigma_{m+1})_{m\in\mathbb{Z}}$.  Associated to each $\omega_0=(\sigma_m)_{m\in\ZM}\in\Omega_0$ one has two sequences
\begin{align*}
 t_{\omega_0}
&\;=\;
\left(t_{\omega_0}(n)\right)_{n\in\mathbb{Z}}
\hspace{0.5mm}\;=\;(\ldots,\baret_{\sigma_0},\baret_{\sigma_1},\ldots)
\;,
\\
 v_{\omega_0}
&\;=\;
\left(v_{\omega_0}(n)\right)_{n\in\mathbb{Z}}
\;=\;(\ldots,\barev_{\sigma_0},\barev_{\sigma_1},\ldots)\;.
\end{align*}
These sequences are not necessarily invariant under shifts of the index because the origin is always a left edge of a polymer. In order to pass into the usual shift invariant framework, one can proceed similarly as in the construction of the Palm distribution. 
Set 
$$
\Omega_{K}\;=\;\{\omega_0\in\Omega_0\,:\,L_{\sigma_0}=K\} \times
\{0,\ldots,K-1\}
\;,
\qquad
\Omega=\bigcup\limits_{K=1}^{L}\Omega_K
\;.
$$
Now the left shift $\tau:\Omega\to\Omega$ is defined by
$$
\tau(\omega_0,{k})
\;=\;
\left\{
\begin{array}{ccc}
(\omega_0,{k}+1) & & \mbox{if } \;{k}< L_{\sigma_0}-1 \mbox{ , } \\
& & \\
(\tau_0\omega_0,0) & & \mbox{if } \;{k}= L_{\sigma_0}-1 \mbox{ , }
\end{array}
\right.
$$
\noindent where $\tau_0$ is the left shift on $\Omega_0$. Now for any
set {$A_{K}\subset\left\{\omega_0\in \Omega_0: L_{\sigma_0}=K\right\}$},
one sets for all $k\in\{0,\ldots, K-1\}$
$$
\PP(A_K\times\{k\})
\;=\;\frac{\PP_0(A_K)}{\langle L_\sigma\rangle}\,
\mbox{ . }
$$
\noindent It
can then be verified that $\PP$ is invariant and ergodic w.r.t. the $\ZM$-action
$\tau$. Finally, for $\omega=(\omega_0,k)$ let us introduce sequences of positive and real numbers respectively by setting
$$
t_\omega(n)
\;=\;
t_{\omega_0}(n+k)\;,
\qquad
v_\omega(n)
\;=\;
v_{\omega_0}(n+k)\;,
\qquad n\in\ZM
\;.
$$
These are the matrix entries of the Jacobi matrix $H_\omega$ which we call the polymer
Hamiltonian of the configuration $\omega$. Namely, it 
is defined by
\begin{equation}
\label{eq-polymerHam}
(H_\omega\psi)(n)
\;=\;
-\,t_{\omega}(n+1) \psi(n+1)
\,+\,v_{\omega}(n)\psi(n)\,-\,t_{\omega}(n)\psi(n-1)
\mbox{ , }
\qquad
\psi\in\ell^2(\ZZ)
\mbox{ , }
\end{equation}
\noindent  and $(H_\omega)_{\omega\in\Omega}$
becomes a family of random operators.
The polymer transfer matrices $T^E_\sigma$ at energy
$E\in\RR$ over a polymer $\sigma=(K,\baret_{\sigma}(0),\ldots,\baret_{\sigma}(K-1),\barev_{\sigma}(0),\ldots,\barev_{\sigma}(K-1))$ are introduced by
\begin{equation}
\label{eq-transfer}
T^E_\sigma
\;=\;
\prod\limits_{k=1}^K
\bareT_{\barev_{{\sigma}}(k-1)-E,\baret_{{\sigma}}(k-1)} 
\mbox{ , }
\qquad \mbox{where }\;\;\;\;\;
\bareT_{\barev,\baret}\;=\;\frac{1}{{\baret}}\,
\left(\begin{array}{cc} {\barev} & {-}\baret^2 \\ 1 & 0 \end{array} \right)
\mbox{ . }
\end{equation}
\noindent The transfer matrices over several polymers are then
\begin{equation}
\label{eq-transsevpol}
T_{\omega_{{0}}}^E(k,m) = T_{\sigma_{k-1}}^E \cdot T_{\sigma_{k-2}}^E \cdot
\ldots \cdot T_{\sigma_m}^E
\mbox{ , }
\qquad k>m 
\mbox{ , } 
\end{equation}

\noindent and $T_{\omega_{{0}}}^E(k,m) = T_{\omega_{{0}}}^E(m,k)^{-1}$ if $k<m$,
$T_{\omega_{{0}}}^E(m,m)={\bf 1}$.

\begin{defini}
\label{def-critical}
An energy $E_c\in\RR$ is called a {\rm hyperbolic critical energy} for the random family
$(H_\omega)_{\omega\in\Omega}$ of polymer Hamiltonians if
the polymer transfer matrices  $T_\sigma^{E_c}$ are hyperbolic {\rm (}i.e.
$|\mbox{\rm Tr}(T_\sigma^{E_c})|>2${\rm )} or equal to $\pm{\bf 1}$ and
commute for all $\sigma,\sigma'\in\Sigma$:
\begin{equation}
\label{eq-critical}
[T^{E_c}_\sigma,T^{E_c}_{\sigma'}]\;=\;0
\;.
\end{equation}
\end{defini}

Note that the critical energies considered in \cite{JSS} were elliptic, namely 
$|\mbox{\rm Tr}(T_\sigma^{E_c})|<2$ or $T_\sigma^{E_c}=\pm{\bf 1}$. The case of parabolic
critical energies was considered in \cite{DKS}.
The definition of the critical energy
assures that there exists a real invertible
matrix $M$ with unit determinant transforming $T^{E_c}_{\sigma}$ for all $\sigma$ 
simultaneously into diagonal hyperbolic matrices:
\begin{equation}
\label{eq-tric}
MT^{E_c}_{\sigma}M^{-1}
\;=\;\pm\,D_{\kappa_{\sigma}}\;,
\qquad
D_{\kappa_{\sigma}}
\;=\;
\left(\begin{array}{cc}
\kappa_{\sigma} & 0 \\ 0 & \frac{1}{\kappa_{\sigma}}
\end{array} \right)
\;,
\end{equation}
where the sign $\pm$ is chosen such that $\kappa_\sigma>0$.
For $\kappa_\sigma\not=1$, the matrix $D_\sigma$ is a hyperbolic matrix from SL$(2,\RM)$ in its usual normal form.

\vspace{.2cm}

\noindent\textbf{Hypothesis:}
{\it The random variable $\kappa_{\sigma}$ satisfies the following:}
$$
\mbox{\rm (i) } \;
\mathbf{P}\left(\kappa_{\sigma}\neq 1\right)>0\;,
\qquad
\mbox{\rm (ii) } \;
\exists\;\nu\neq 0\;:\;\langle \kappa^{\nu}_{\sigma}\rangle\,=\,1
\;.
$$

\noindent\textbf{Remark.} Items (i) and (ii) imply that the support of $\kappa_{\sigma}$ intersects both $(0,1)$ and $(1,\infty)$ non-trivially. Let us also note that the strict convexity of $\xi\mapsto \kappa^{\xi}$ implies the uniqueness of~$\nu$. If $\nu>0$, then by Jensen's~inequality,
$$
\partial_\xi\,
\langle \kappa^{\xi}_{\sigma}\rangle\,\mid_{\xi=0}
\;=\;
\langle \log\kappa_{\sigma}\rangle 
\;=\;
\nu^{-1}\langle \log\kappa_{\sigma}^{\nu}\rangle
\;<\; 
\nu^{-1}\log\langle \kappa_{\sigma}^{\nu}\rangle
\;=\;0
\;,
$$
with possibly $\langle \log\kappa_{\sigma}\rangle =-\infty$. If $\nu<0$, then $\langle \log\kappa_{\sigma}\rangle>0$. In the following we may assume that $\nu>0$, as otherwise $(\kappa_{\sigma},M)$ can be replaced by $(\kappa_{\sigma}^{-1},IM)$, where $\imath I$ is the second Pauli matrix. 
\hfill $\diamond$

\vspace{.2cm}

\noindent {\bf Example} Let $L_\sigma=2$ and ${\barev}_\sigma=(0,0)$ and ${\baret}_\sigma=(\hat{t}_\sigma(0),1)$. Then 
$$
T^E_\sigma
\;=\;
\bareT_{{-E,1}}\bareT_{{-E,\hat{t}_\sigma(0)}}
\;=\;
\begin{pmatrix}
\frac{E^2-1}{\hat{t}_\sigma(0)} & E\hat{t}_\sigma(0) \\
\frac{-E}{\hat{t}_\sigma(0)} & -\hat{t}_\sigma(0)
\end{pmatrix}
\;=\;
- \left[
\begin{pmatrix}
1 & -E \\
E & 1
\end{pmatrix}
\,+\,\Oo(E^2)
\right]
\begin{pmatrix}
\frac{1}{\hat{t}_\sigma(0)} & 0 \\
0 & \hat{t}_\sigma(0)
\end{pmatrix}
\;.
$$
Hence $E_c=0$ is a hyperbolic critical energy and the basis transformation $M$ in \eqref{eq-tric} is the identity. Note that the first factor on the r.h.s. is to lowest order in $E$ a rotation by $E$.
\hfill $\diamond$

\vspace{.2cm}

It will be convenient to always expand the polymer transfer matrix around the critical energy similar as in the example. More precisely, let us introce real numbers $a_\sigma,b_\sigma,c_\sigma$ by
\begin{equation}
\label{eq-Expand}
MT^{E_c+\epsilon}_{\sigma}M^{-1}
\;=\;
\pm
\left[
\one\,+\,a_\sigma\epsilon
\begin{pmatrix}
0 & -1 \\
1 & 0
\end{pmatrix}
\,+\,b_\sigma\epsilon
\begin{pmatrix}
0 & 1 \\
1 & 0
\end{pmatrix}
\,+\,c_\sigma\epsilon
\begin{pmatrix}
1 & 0 \\
0 & -1
\end{pmatrix}
\,+\,\Oo(\epsilon^2)
\right]
D_{\kappa_\sigma}
\;.
\end{equation}
In the above example, one has $E_c=0$ and $b_\sigma=c_\sigma=0$ and $a_\sigma=1$. In general:

\begin{proposi}
\label{prop-CoeffProperties}
The inequalities $a_{\sigma}\geq 0$ and $a^2_\sigma\geq b_\sigma^2+c_\sigma^2$ hold for all $\sigma\in\Sigma$.
\end{proposi}

\noindent {\bf Proof.}
Let us set $J=\binom{0\;-1}{1\;\;\;0}$ and recall that
\begin{align*}
& 
\big(MT^{E}_{\sigma}M^{-1}\big)^*J^*\,\partial_E \big(MT^{E}_{\sigma}M^{-1}\big)
\\
&
\;\;\;=\;
(T^{E}_{\sigma}M^{-1})^* 
J^*
\sum_{k=0}^{L_\sigma -1}
\Big[\prod\limits_{{l=k+1}}^{{L_{\sigma}-1}}
\bareT_{\barev_{{\sigma}}(l)-E,\baret_{{\sigma}}(l)} 
\Big]
\begin{pmatrix}
\tfrac{-1}{\baret_{{\sigma}}(k)} & 0 \\ 0 & 0
\end{pmatrix}
\Big[
\prod\limits_{{l=0}}^{{k-1}}
\bareT_{\barev_{{\sigma}}(l)-E,\baret_{{\sigma}}(l)} 
\Big]
M^{-1}
\\
&
\;\;\;=\;
\sum_{k=0}^{L_\sigma -1}
(M^{-1})^* 
\Big[
\prod\limits_{{l=0}}^{{k}}
\bareT_{\barev_{{\sigma}}(l)-E,\baret_{{\sigma}}(l)} 
\Big]^*J^*
\begin{pmatrix}
\tfrac{-1}{\baret_{{\sigma}}(k)} & 0 \\ 0 & 0
\end{pmatrix}
\Big[
\prod\limits_{{l=0}}^{{k-1}}
\bareT_{\barev_{{\sigma}}(l)-E,\baret_{{\sigma}}(l)} 
\Big]
M^{-1}
\\
&
\;\;\;=\;
\sum_{k=0}^{L_\sigma -1}
(M^{-1})^* 
\Big[
\prod\limits_{{l=0}}^{{k-1}}
\bareT_{\barev_{{\sigma}}(l)-E,\baret_{{\sigma}}(l)} 
\Big]^*{\big(\bareT_{\barev_{{\sigma}}(k)-E,\baret_{{\sigma}}(k)}\big)^{*}}J^*
\begin{pmatrix}
\tfrac{-1}{\baret_{{\sigma}}(k)} & 0 \\ 0 & 0
\end{pmatrix}
\Big[
\prod\limits_{{l=0}}^{{k-1}}
\bareT_{\barev_{{\sigma}}(l)-E,\baret_{{\sigma}}(l)} 
\Big]
M^{-1}
\\
&
\;\;\;{=}\;
\sum_{k=0}^{L_\sigma -1}
(M^{-1})^* 
\Big[
\prod\limits_{{l=0}}^{{k-1}}
\bareT_{\barev_{{\sigma}}(l)-E,\baret_{{\sigma}}(l)} 
\Big]^*
\begin{pmatrix}
\tfrac{1}{\baret^2_{{\sigma}}(k)} & 0 \\ 0 & 0
\end{pmatrix}
\Big[
\prod\limits_{{l=0}}^{{k-1}}
\bareT_{\barev_{{\sigma}}(l)-E,\baret_{{\sigma}}(l)} 
\Big]
M^{-1}
\;.
\end{align*}
Now this matrix is manifestly non-negative. On the other hand, replacing \eqref{eq-Expand} gives
\begin{align*}
\big(MT^{E_c}_{\sigma}M^{-1}\big)^*J^*\partial_E \big(MT^{E_c}_{\sigma}M^{-1}\big)
&
\;=\;
D_{\kappa_\sigma}^*J^*
\left[
a_\sigma
\begin{pmatrix}
0 & -1 \\
1 & 0
\end{pmatrix}
\,+\,b_\sigma
\begin{pmatrix}
0 & 1 \\
1 & 0
\end{pmatrix}
\,+\,c_\sigma
\begin{pmatrix}
1 & 0 \\
0 & -1
\end{pmatrix}
\right]
D_{\kappa_\sigma}
\\
&
\;=\;
D_{\kappa_\sigma}^*
\left[
a_\sigma
\begin{pmatrix}
1 & 0 \\
0 & 1
\end{pmatrix}
\,+\,b_\sigma
\begin{pmatrix}
1 & 0 \\
0 & -1
\end{pmatrix}
\,-\,c_\sigma
\begin{pmatrix}
0 & 1 \\
1 & 0
\end{pmatrix}
\right]
D_{\kappa_\sigma}
\;.
\end{align*}
Non-negativity of this expression implies the claim.
\hfill $\Box$

\subsection{Pr{\"u}fer variables}
\label{sec-Pruefer}

This section briefly recalls definitions and basic properties of the free Pr\"ufer variables and $M$-modified Pr\"ufer variables. As this can be spelled out for every single realization $\omega$, the index is dropped. 
Let $(t(n))_{n\in\ZZ}$ be a sequence of positive numbers and
$(v(n))_{n\in\ZZ}$ a sequence of real numbers. As in
(\ref{eq-polymerHam}) they define a Jacobi matrix $H$. Given an initial
phase $\theta(0)\in\RR$ and an energy $E\in\RM$, let us construct the
formal solution $(u^E(n))_{n\in\ZZ}$ by
\begin{equation}
\label{eq-eigenstates}
-t(n+1)u^E(n+1)+v(n)u^E(n)-t(n)u^E(n-1)
\;=\;
E u^E(n)
\mbox{ , }
\end{equation}
\noindent and the initial conditions
$$
\left(\begin{array}{c} t(0)\,u^E(0) \\ u^E(-1)
\end{array} \right)
\;=\;
\left(\begin{array}{c} \cos(\theta(0)) \\ \sin(\theta(0))
\end{array} \right)
\mbox{ . }
$$
\noindent Using the definition (\ref{eq-transfer}) of the single site
transfer matrices $\bareT_{{\barev,\baret}}$, the transfer matrix from site $k$ to $n$
is introduced by
$$
\bareT^E(n,k)\;=\;\prod_{l={k}}^{{n-1}}\,\bareT_{{v(l)-E},t(l)}
\mbox{ . }
$$
\noindent It allows to rewrite the (formal) eigenfunction equation  
(\ref{eq-eigenstates}) as
\begin{equation}
\label{eq-transferrel}
\left(\begin{array}{c} t(n)\,u^E(n) \\ u^E(n-1)
\end{array} \right)
\;=\;
\bareT^E(n,k)\;
\left(\begin{array}{c}t(k)\, u^E(k) \\ u^E(k-1)
\end{array} \right)
\mbox{ . }
\end{equation}
The free Pr{\"u}fer phases $\theta^{0,E}(n)$ and amplitudes
$R^{0,E}(n)>0$ are now defined by
\begin{equation}
\label{eq-freeprufer}
R^{0,E}(n) \,
\left( \begin{array}{c} \cos (\theta^{0,E}(n))
\\ \sin(\theta^{0,E}(n)) \end{array} \right)
\;=\;
\left( \begin{array}{c} t(n)
u^E(n) \\ u^E(n-1) \end{array} \right)
\mbox{ , }
\end{equation}
\noindent the above initial conditions as well as 
$$
-\frac{\pi}{2}  < \theta^{0,E}(n+1) - \theta^{0,E}(n)
< \frac{3\pi}{2}
\mbox{ . }
$$
\noindent Note that the dependence of the Pr{\"u}fer
variables on $\theta(0)$ is suppressed. Recall that
$\partial_E \theta^{0,E}(n)$ is strictly positive for $n\ge
2$ and strictly negative for $n\le -2$ ({\it e.g.}~\cite{JSS}, Lemma 2).

\vspace{.2cm}

Let $\Pi_N$ be the projection on $\ell^2(\{0,\ldots,N-1\})$ and
denote the associated finite-size Jacobi matrix by $H_N=\Pi_N H\Pi_N$.
As $H_N$ has Dirichlet boundary conditions, let us choose
$u^E(-1)=0$ and $t(0)u^E(0)=1$ as initial conditions in the recurrence
relation (\ref{eq-eigenstates}). This corresponds to an initial
Pr{\"u}fer phase $\theta(0)=0$. 
The oscillation theorem ({\it e.g.} \cite{JSS}) implies 
\begin{equation}
\label{eq-oscithm}
\left| \frac{1}{\pi}\;\theta^{0,E}(N) \;-\;
\#\;\left\{\mbox{negative eigenvalues of}\;\;(H_N-E)\,
\right\}
\right|
\; \leq\;
\frac{1}{2}
\mbox{ . }
\end{equation}
Next let us pass to $M$-modified Pr\"ufer variables. Hence fix $M\in \mbox{\rm SL}(2,\RR)$.  Define a
smooth function $m:\RR\to\RR$ with 
$m(\theta+\pi)=m(\theta)+\pi$ and
$0 < C_1 \le  m' \le C_2 < \infty$, by
$$
r(\theta)e_{m(\theta)}=Me_{\theta},
\qquad
r(\theta)>0
\mbox{ , }
\qquad
m(0)\in[-\pi,\pi)
\;,
$$
\noindent where $e_\theta\in\RM^2$ is the unit vector as defined in the introduction. Then the $M$-modified Pr{\"u}fer variables
$(R^{M,E}(n),\theta^{M,E}(n))\in \RR_+\times\RR$ for the initial condition
$\theta^{M,E}(0)=\theta=m(\theta^{0})$
are given by
\begin{equation}
\label{eq-prufer1}
\theta^{M,E}(n)\;=\; m(\theta^{0,E}(n)) \mbox{ , }
\end{equation}
\noindent and 
\begin{equation}
\label{eq-prufer2}
{R^{M,E}(n)}\left(\begin{array}{c} \cos
(\theta^{M,E}(n))
\\
 \sin(\theta^{M,E}(n))
\end{array}
\right)
\;=\;
M \left( \begin{array}{c} t(n)\, u^E(n) \\ u^E(n-1)
\end{array} \right) 
\mbox{ , }
\end{equation}

\noindent where the dependence on the initial phase is again
suppressed. 
Then \eqref{eq-oscithm} implies \cite{JSS}
\begin{equation}
\label{eq-oscithm2}
\left| \frac{1}{\pi}\;\theta^{M,E}(N) \;-\;
\#\;\{\mbox{negative eigenvalues of }\;\;(H_N-E)\,\} \right| \;\leq\;
\frac{5}{2}
\mbox{ . }
\end{equation}

\subsection{Covariant Jacobi matrices}
\label{sec-covJac}

Let $(\Omega,\tau,\ZZ,\PP)$ be a compact space $\Omega$, endowed with a
$\ZZ$-action $\tau$ and a $\tau$-invariant and ergodic probability measure
$\PP$. For a function $f\in
L^1(\Omega,\PP)$, let us denote $\EE(f(\omega))=\int
d\PP(\omega)\,f(\omega)$.
A strongly continuous family
$(H_\omega)_{\omega\in\Omega}$ of two-sided tridiagonal, self-adjoint
matrices on
$\ell^2(\ZZ)$ is
called covariant if the covariance relation
$UH_\omega U^*=H_{\tau\omega}$ holds where $U$ is the translation on
$\ell^2(\ZZ)$.
$H_\omega$ is characterized by two sequences $(t_\omega(n))_{n\in\ZZ}$
and $(v_\omega(n))_{n\in\ZZ}$ such that
(\ref{eq-polymerHam}) holds.

\vspace{.2cm}

The IDS at energy $E\in\RR$ of the family
$(H_\omega)_{\omega\in\Omega}$
can $\PP$-almost surely be defined by \cite{PF,BL,AW}
\begin{equation}
\label{eq-DOS}
\Nn(E)
\;=\;
\lim_{N\to\infty}\;\frac{1}{N}
\;\mbox{Tr}(\chi_{(-\infty,E]}(\Pi_NH_\omega\Pi_N))
\mbox{ , }
\end{equation}
\noindent while the Lyapunov exponent $\gamma(E)$ for $E\in\RR$
is $\PP$-almost surely given by the formula
$$
\gamma(E)
\;=\;
\lim_{N\to\infty}\frac{1}{N}
\log\left(\left\|{\bareT}^E_\omega(N,0)
\right\|\right)
\mbox{ , }
$$
\noindent where the transfer matrix $\bareT^E_{\omega}(N,0)$ from
site $0$ to $N$ is defined as in Section~\ref{sec-Pruefer}. Both the 
IDS and the Lyapunov
exponent are self-averaging quantities, notably
an average over $\PP$ may be introduced before taking the
limit without changing the result \cite{PF}. The IDS and the Lyapunov exponent are
linked by the Thouless formula (see~\cite{CL}, p.~376)
\begin{align}\label{Thouless_formula}
\gamma(E)
\;=\;
-\,\left\langle\log( t(0))\right\rangle\;+\;\int \Nn(dE')\,\log(|E-E'|)
\;,
\qquad
E\in\CM
\;.
\end{align}

\vspace{.2cm}

For each $H_\omega$ let
$(R^{M,E}_\omega(n),\theta^{M,E}_\omega(n))$ denote the associated
$M$-modified Pr{\"u}fer variables with some initial condition, 
then according to (\ref{eq-oscithm2})
\begin{equation}
\label{eq-IDS}
\Nn(E)
\;=\;
\lim_{N\to\infty}\;\frac{1}{\pi}\,\frac{1}{N}\left\langle\theta^{M,E}_\omega(N)\right\rangle
\mbox{ . }
\end{equation}
The r.h.s. is the rotation number and the equality \eqref{eq-IDS} expresses what is called the rotation number calculation of the IDS.

\subsection{Modified polymer Pr\"ufer variables}
\label{sec-ModPolPruef}

While the exposition in the last two sections was generic, we now specify to the random polymer model with a hyperbolic critical energy $E_c$. Then there is a naturally associated basis change $M$ such that
the transfer matrices over a polymer $\sigma\in\Sigma$ are given by \eqref{eq-Expand}. It is now natural to consider the $M$-modified Pr\"ufer variables $\theta^{M,E}_\omega(m)$ not on all sites of $m\in\ZM$, but rather only on the left boundaries of the $n$th polymer which for a configuration $\omega=\big((\sigma_m)_{m\in\ZM},k)$ is given by $k+\sum_{l=0}^{n-1}L_{\sigma_l}$. Hence let us introduce the $M$-modified polymer Pr\"ufer variables by
\begin{equation}
\label{eq-PolymerPruefer}
\theta^{\epsilon}_\omega(n)
\;=\;
\theta^{M,E_c+\epsilon}_\omega\big(k+\sum_{l=0}^{n-1}L_{\sigma_l}\big) \;\mod\;\pi\;,
\end{equation}
together with a suitable choice of lift that will be fixed next. For that purpose, let us recall that by the elementary gap labelling of the gap at $E_c$ for the periodic operator given by periodizing the polymer block $\sigma$,
there exists an integer $l_\sigma\in\{0,\ldots,L_\sigma\}$ such that 
\begin{equation}
\label{eq-PolymerPruefer0}
\theta^{0,E_c}_{\omega}(L_{\sigma}-k)\,-\,\theta^{0,E_c}_{\omega}(-k)
\;=\;
\pi\,l_\sigma
\end{equation}
where $\omega=(\omega_0,k)$ is such that $\omega_0=(\sigma_n)_{n\in\ZM}$ with $\sigma_0=\sigma$. Then the IDS of the random polymer Hamiltonian $(H_\omega)_{\omega\in\Omega}$ at the critical energy is given by
$$
\Nn(E_c)
\;=\;
\frac{\langle l_\sigma\rangle}{\langle L_\sigma\rangle}
\;.
$$
Then \eqref{eq-PolymerPruefer0} implies that
$$
\theta^{M,E_c}_{\omega}(L_{\sigma}-k)\,-\,\theta^{M,E_c}_{\omega}(-k)
\;=\;
\pi\,l_\sigma
\;,
$$
where still $\omega=(\omega_0,k)$ is such that $\omega_0=(\sigma_n)_{n\in\ZM}$ with $\sigma_0=\sigma$. Then the lift in \eqref{eq-PolymerPruefer} is fixed by
$$
\theta^{\epsilon}_\omega(n)-\theta^{\epsilon}_\omega(n-1)
\;=\;
\theta^{M,E_c+\epsilon}_\omega\big({-}k+\sum_{l=0}^{n-1}L_{\sigma_l}\big)
\,-\,\theta^{M,E_c+\epsilon}_\omega\big({-}k+\sum_{l=0}^{n-2}L_{\sigma_l}\big) 
\,-\,\pi l_{\sigma_n}
\;.
$$
Consequently, by iterating this and taking in \eqref{eq-IDS} subsequences only on the polymer boundaries,
\begin{align}
\Nn(E_c+\epsilon)
& 
\;=\;
\frac{1}{\pi}\,\lim_{N\to\infty}\,
\frac{1}{\sum_{n={0}}^{N{-1}}L_{\sigma_n}}\;
\Big\langle \theta^{\epsilon}_\omega(N)\;+\;\pi \sum_{n=1}^Nl_{\sigma_n}\Big\rangle
\nonumber
\\
&\;=\;
\Nn(E_c)
\;+\;
\frac{1}{\pi\,\langle L_\sigma\rangle}\,\lim_{N\to\infty}\;\frac{1}{N}\left\langle\theta^{\epsilon}_\omega(N)\right\rangle
\;.
\label{eq-IDSRotnumber}
\end{align}

Due to the set-up, the $M$-modified polymer Pr\"ufer variables satisfy
\begin{equation}
\label{eq-ModPruefDyn}
{R}^{\epsilon}_{\omega}(n)\begin{pmatrix}\cos(\theta_{\omega}^{\epsilon}(n))\\ \sin(\theta_{\omega}^{\epsilon}(n))\end{pmatrix}
\;=\;
MT^{E_c+\epsilon}_{\sigma_n}M^{-1}\,
\begin{pmatrix}\cos(\theta_{\omega}^{\epsilon}(n-1))\\ \sin(\theta_{\omega}^{\epsilon}(n-1))\end{pmatrix}\;,
\end{equation}
where ${R}^{\epsilon}_{\omega}(n)>0$ is a normalization factor that is irrelevant for the present purposes. One can now replace \eqref{eq-Expand} for $MT^{E_c+\epsilon}_{\sigma}M^{-1}$. It is, however, useful to include the term resulting from $c_\sigma$ into the hyperbolic factor. The cost is a commutator of higher order $\epsilon^2$. Hence let us introduce the notations
\begin{equation}
\label{eq-ModPruefDyn2}
MT^{E_c+\epsilon}_{\sigma}M^{-1}
\;=\;
Q^\epsilon_{\sigma}\,D_{\kappa_{\sigma}(1+\epsilon c_\sigma)}
\;,
\end{equation}
with
$$
Q^{\epsilon}_{\sigma}
\;=\;
\one\,+\,a_\sigma\epsilon
\begin{pmatrix}
0 & -1 \\
1 & 0
\end{pmatrix}
\,+\,b_\sigma\epsilon
\begin{pmatrix}
0 & 1 \\
1 & 0
\end{pmatrix}
\,+\,
\epsilon^2 \,A^\epsilon_\sigma\;,
\qquad
A_{\sigma}^{\epsilon}
\;=\;
\begin{pmatrix}
\alpha_{\sigma}^{\epsilon} & \beta_{\sigma}^{\epsilon} \\ \gamma_{\sigma}^{\epsilon} & \delta_{\sigma}^{\epsilon}
\end{pmatrix}
\;.
$$
Modifying $\kappa_\sigma$ to $\kappa_{\sigma}(1+\epsilon c_\sigma)$ is, for $\epsilon$ sufficiently small, not of any relevance, but does lead to heavier notations and some inessential complications in the argument below, so we simply suppose $c_\sigma=0$ for all $\sigma\in\Sigma$. Note that this is the case anyhow in the random hopping model, {\it cf.} \eqref{eq-GenDyn}. Of importance will be, however, to make some assumptions on the random coefficients of $Q^{\epsilon}_{\sigma}$. We will assume that the following are positive and finite quantities:
\begin{align}
\label{assumption_on_a_and_b}
& 
C_1
\;=\;
\essinf\left(a_{\sigma}-|b_{\sigma}|\right)
\;,
\quad
C_2
\;=\;
\esssup\big(a_{\sigma}+|b_{\sigma}|\big)
\;,
\quad
C_3
\;=\;
\sup\limits_{|\epsilon|\leq 1}\esssup\|A_{\sigma}^{\epsilon}\|
\;,
\end{align}
where the essential infimum and supremum are taken over $\sigma\in\Sigma$. Even though it can be worked around it (see Section~\ref{sec-unbounded}), the arguments below become simpler when we also assume finiteness of 
$$
C_4\;=\;\esssup\kappa_\sigma
\;.
$$

\noindent {\bf Example} These assumptions are satisfied in case of~\eqref{eq-GenDyn} provided the support of $t(2n+1)$ is compact in $(0,\infty)$. Indeed, then $C_3=\esssup t(2n+1)^{-2}<\infty$ and 
$$
a_{\sigma}
\;=\;
\frac{t(2n+1)^{-2}+1}{2}\;,
\qquad 
b_{\sigma}
\;=\;
\frac{t(2n+1)^{-2}-1}{2}\;,
\qquad 
c_{\sigma}\;=\;0
\;,
$$
so that $C_1=\min\{1,\essinf t(2n+1)^{-2}\}>0$ and $C_2=\max\{1,\esssup t(2n+1)^{-2}\}<\infty$.
\hfill $\diamond$

\section{Bound on the rotation number}
\label{sec-MainTechnics}

\begin{figure}
\centering
\includegraphics[width=16cm]{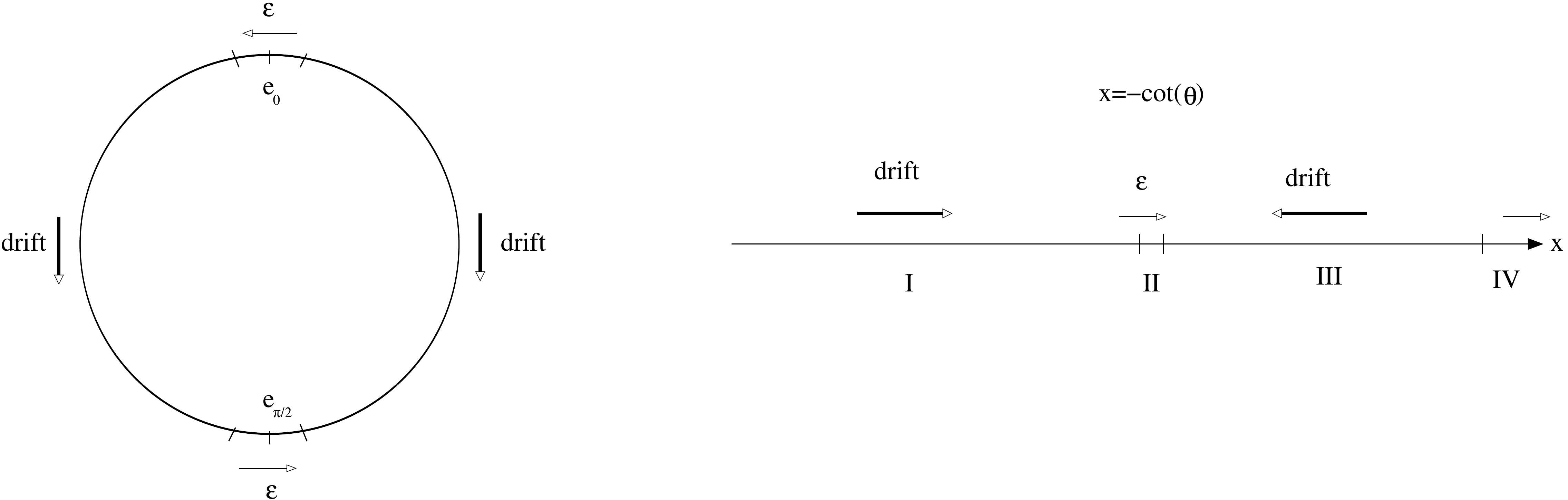} 
\caption{Schematic representation of the random dynamics of the Dyson-Schmidt variables and the region described in the text.}
\label{fig-DysonSchmidt}
\end{figure}

In this section, we prove an upper bound of the {average} rotation number on the r.h.s. of \eqref{eq-IDSRotnumber} in the vicinity of a critical energy $E_c$. This will be based on a detailed analysis of the modified polymer Pr\"ufer phases defined in Section~\ref{sec-ModPolPruef} and, in particular, a probabilistic control on the average time to make a loop in projective space. It will be convenient to achieve this for the Dyson-Schmidt variables defined by
\begin{equation}
\label{eq-DysSch}
x_{\omega}^{\epsilon}(n)
\;=\;-\cot(\theta_{\omega}^{\epsilon}(n))
\;.
\end{equation}
The map $\theta\in(-\frac{\pi}{2},\frac{\pi}{2}]\mapsto x=-\cot(\theta)\in\overline{\RM}$ is an orientation preserving bijection onto the one-point compactification $\overline{\RM}=\RM\cup\{\infty\}$ which is also called the stereographic projection (some other authors do not include the sign or use the tangent). Just as in \eqref{eq-PrueferDyn}, the dynamics of the $x_{\omega}^{\epsilon}(n)$ as deduced from \eqref{eq-ModPruefDyn} is given by the M\"obius transformation with the matrix given in \eqref{eq-ModPruefDyn2}. The M\"obius action of a $2\times 2$ matrix $M$ on $\RM$ is denoted by $M\cdot x$. As this dynamics is generated by two consecutive M\"obius actions by  $D_{\kappa_{\sigma_n}}$ and $Q^\epsilon_{\sigma_n}$ (recall that we suppose $c_\sigma=0$), it is useful to set
$$
x_{\omega}^{\epsilon}(n-\tfrac{1}{2})
\;=\;
D_{\kappa_{\sigma_n}}\cdot
x_{\omega}^{\epsilon}(n-1)
\;,
$$
so that
$$
x_{\omega}^{\epsilon}(n)
\;=\;
Q^\epsilon_{\sigma_n}\cdot
x_{\omega}^{\epsilon}(n-\tfrac{1}{2})
\;=\;
(Q^\epsilon_{\sigma_n}D_{\kappa_{\sigma_n}})
\cdot
x_{\omega}^{\epsilon}(n-1)
\;.
$$
The dynamics is shown in Fig.~\ref{fig-DysonSchmidt}.

\subsection{Deterministic bounds on Dyson-Schmidt variables}
\label{sec-Deterministic}

The first step of the analysis consists of deterministic arguments to verify that the scenario sketched in the introduction is valid. Hence let us drop all indices on $Q_{\sigma_n}^{\epsilon}$, $D_{\kappa_{\sigma_n}}$, $x_{\omega}^{\epsilon}(n)$, $a_{\sigma_n}$, $b_{\sigma_n}$, $\alpha_{\sigma_n}^{\epsilon}$, $\beta^{\epsilon}_{\sigma_n}$, $\gamma^{\epsilon}_{\sigma_n}$ and $\delta^{\epsilon}_{\sigma_n}$ in order to improve readability. Furthermore, let us spell out the action of $Q$ and $D$ on $x$ explicitly:
\begin{equation}
\label{eq-DynExplicit}
Q\cdot x\;=\;\frac{(1+\epsilon^2\alpha)x+(a{-b}-\epsilon\beta)\epsilon}{1+\epsilon^2\delta-(a{+b}+\epsilon\gamma)\epsilon x}
\;,
\qquad
D\cdot x\;=\;\kappa^2x
\;.
\end{equation}
As the effect of $Q\cdot$ and $D\cdot$ is strongly dependent on $x\in\overline{\RM}$, it will be useful to split the compactified real line in several regions. This splitting will depend on a parameter $\kConst>1$ associated to which we also fix $\KConst =2C_2(1-\kConst^{-1})^{-1}$. Then set:
\begin{align*}
\mathfrak{R}_{\rm I}
&
\,=\,(-\infty,0]\;,
\\
\mathfrak{R}_{\rm II}
&
\,=\,
\left\{x\in\mathbb{R}\;:\; 0<x<\KConst \epsilon\right\}\;,
\\
\mathfrak{R}_{\rm III}
&
\,=\,
\left\{x\in\mathbb{R}\;:\; \KConst \epsilon\leq x \leq (\KConst {\epsilon})^{-1} \right\}
\;,
\\
 \mathfrak{R}_{\rm IV}
&
\,=\,
\left\{x\in\mathbb{R}\cup\{\infty\}\;:\;x>(\KConst \epsilon)^{-1}\right\}
\;.
\end{align*}

\begin{lemma}\label{maximum_factor}
There exists an ${\epsilon}_0={\epsilon}_0(\kConst,C_1,C_3)\in (0,1)$ such that all $\epsilon\in (0,{\epsilon}_0]$ satisfy
\begin{align}
\label{maximum_factor_statement_1}
& \KConst \epsilon\leq x\leq (\KConst \epsilon)^{-1}\qquad \quad  \Longrightarrow \qquad Q\cdot x\leq \kConst x\,,
& (\mathfrak{R}_{\rm III})
\\
\label{maximum_factor_statement_2}
&
\KConst \epsilon\leq x\leq (\KConst \epsilon)^{-1}\qquad\quad   \Longrightarrow \qquad (DQ)\cdot x\leq \kConst\kappa^2x\,,
& (\mathfrak{R}_{\rm III})
\\
\label{maximum_factor_statement_3}
& x> 0 \mbox{ and } (QD)\cdot x\leq 0\;\, \hspace{-0.1mm} \Longrightarrow\qquad D\cdot x > (\KConst \epsilon)^{-1}\;,
& (\mbox{into } \mathfrak{R}_{\rm I}\mbox{ via } \mathfrak{R}_{\rm IV})
\\
\label{maximum_factor_statement_4}
& x\leq 0\qquad \qquad \qquad \qquad \hspace{0.6mm} \Longrightarrow\qquad Q\cdot x<\KConst \epsilon\;.
& (\mbox{from } \mathfrak{R}_{\rm I})
\end{align}
\end{lemma}

\noindent\textbf{Proof.} Let us first note that $Q$ is a (random) rotation by terms of order $\epsilon$, so that there is no real solution of the fixed point equation $Q\cdot x=x$. However, for $\kConst >1$, there are two real roots of the quadratic equation $Q\cdot x=\kConst x$ which are given by
$$
x_{\pm}
\;=\;
\frac{\kConst(1+\epsilon^2\delta)-(1+\epsilon^2\alpha)}{2\kConst\epsilon(a{+b}+\epsilon\gamma)}
\left(1\;\pm\;\sqrt{1-\frac{4\kConst\epsilon^2(a{+b}+\epsilon\gamma)(a{-b}-\epsilon\beta)}{[\kConst(1+\epsilon^2\delta)-(1+\epsilon^2\alpha)]^2}}\textnormal{ }\right)
\;.
$$
%
%
Indeed, the term under the square root is positive by  the assumptions $C_1>0$ and $C_2<\infty$ for $\epsilon$ sufficiently small. For $y\in[0,1]$ one has $1-y\leq \sqrt{1-y}$. With this at hand, one readily checks that $[y_-,y_+]\subset [x_-,x_+]$ for
$$
y_-
\;=\;
\frac{2\epsilon (a{-b}-\epsilon\beta)}{\kConst (1+\epsilon^2\delta)-(1+\epsilon^2\alpha)}
\;,
\qquad 
y_+
\;=\;
\frac{\kConst (1+\epsilon^2\delta)-(1+\epsilon^2\alpha)}{\kConst \epsilon (a{+b}+\epsilon\gamma)}
\;-\;
\frac{2\epsilon (a{-b}-\epsilon\beta)}{\kConst (1+\epsilon^2\gamma)-(1+\epsilon^2\alpha)}
\;.
$$
Hence for $x\in [y_-,y_+]$, one has $Q\cdot x\leq \kConst x$. In the final step of the proof of \eqref{maximum_factor_statement_1}, one now has to check that $[\epsilon \KConst,\frac{1}{\epsilon\KConst}]\subset [y_-,y_+]$. Indeed, using the assumed bounds on the constants in \eqref{assumption_on_a_and_b}, one has (uniformly in $\sigma$) for $\epsilon$ sufficiently small
$$
y_-
\;\leq\;
\frac{{2}\epsilon(C_2+\epsilon C_3)}{\kConst (1-\epsilon^2C_3)-(1+\epsilon^2C_3)}
\;\leq\;
\epsilon\;\frac{2{k}C_2}{k-1}
\;{=}\;
\epsilon\,\KConst
\;,
$$
and
$$
y_+
\;\geq\;
\frac{\kConst (1-\epsilon^2C_3)-(1+\epsilon^2C_3)}{\kConst \epsilon (C_2+\epsilon C_3)}
\;-\;
\frac{2\epsilon(C_2+\epsilon C_3)}{\kConst (1-\epsilon^2C_3)-(1+\epsilon^2C_3)}
\;\geq \;
\frac{1}{\epsilon}\,\frac{\kConst-1}{2\kConst C_2}
\;= \;
\frac{1}{\epsilon\KConst}
\;.
$$
Now the proof of \eqref{maximum_factor_statement_1} is completed. That of \eqref{maximum_factor_statement_2} then follows directly from \eqref{eq-DynExplicit}.

\vspace{.1cm}

As for the proof of~\eqref{maximum_factor_statement_3}, recall that the denominator of $Q\cdot x$ in \eqref{eq-DynExplicit} is positive whenever $0< x\leq (\KConst\epsilon)^{-1}$ (see above). Moreover, the numerator is bounded below $C_1\epsilon(1+C_3)^{-1}>0$ for $\epsilon$ sufficiently small. Thus $Q\cdot x>0$. As $D\cdot $ preserves the sign, the negation of~\eqref{maximum_factor_statement_3} is falsified by replacing $x$ by $D\cdot x$.

\vspace{.1cm}

As for the proof of~\eqref{maximum_factor_statement_4}, it is sufficient to consider the case $Q\cdot x>0$ in which both numerator and denominator in \eqref{eq-DynExplicit} are positive. Using $x<0$, one can now estimate as~follows:
\begin{align*}
Q\cdot x
\;\leq\;
\frac{(C_2+\epsilon C_3)\epsilon}{1-\epsilon^2 C_3}
\;\leq\;
\frac{2C_2\epsilon}{1-\epsilon^2 C_3}
\;=\;
\frac{(\KConst-2C_2(\kConst -1)^{-1})\epsilon}{1-\epsilon^2 C_3}
\;<\;
\epsilon\KConst\;,
\end{align*}
again for sufficiently small $\epsilon$.
\hfill $\square$

\vspace{.2cm}

Let us now collect a few first implications of Lemma~\ref{maximum_factor}. For this purpose, let us use the notation
$$
\mathfrak{R}_{\rm I}^c\;=\;(0,\infty]\;.
$$
Loops on projective space require passages from $\mathfrak{R}_{\rm I}$ to $\mathfrak{R}_{\rm I}^c$ and back to $\mathfrak{R}_{\rm I}$. Since $D\cdot$ preserves the sign, leaving one of the half-lines and entering the other one, is only possible as a consequence of the action $Q\cdot $, namely
\begin{align}
\label{consequence_of_invariance_of_sign_2}
&(x_{\omega}^{\epsilon}(n-1),x_{\omega}^{\epsilon}(n))\;\in\;
\mathfrak{R}_{\rm I}^c\times \mathfrak{R}_{\rm I}
\qquad\Longrightarrow\qquad x_{\omega}^{\epsilon}(n-\tfrac{1}{2})\;\in\;\mathfrak{R}_{\rm I}^c
\;,
\\
\label{consequence_of_invariance_of_sign_1}
&
(x_{\omega}^{\epsilon}(n-1),x_{\omega}^{\epsilon}(n))\;\in\;
\mathfrak{R}_{\rm I}\times \mathfrak{R}_{\rm I}^c
\qquad\Longrightarrow\qquad x_{\omega}^{\epsilon}(n-\tfrac{1}{2})\;\in\;\mathfrak{R}_{\rm I}\;.
\end{align}
Now,~\eqref{consequence_of_invariance_of_sign_2} can be improved, namely by using Lemma~\ref{maximum_factor}, whose penultimate statement~\eqref{maximum_factor_statement_3} implies that $\mathfrak{R}_{\rm I}$ can only be entered by leaving $\mathfrak{R}_{\rm IV}$, {\it i.e.},
\begin{align}\label{leave_condition}
(x_{\omega}^{\epsilon}(n-1),x_{\omega}^{\epsilon}(n))
\;\in\;
\mathfrak{R}_{\rm I}^c\times \mathfrak{R}_{\rm I}
\qquad\Longrightarrow\qquad x_{\omega}^{\epsilon}(n-\tfrac{1}{2})\in\mathfrak{R}_{\rm IV}\,.
\end{align}
Statement~\eqref{consequence_of_invariance_of_sign_1} is can be improved in a similar way. However, it is more useful to understand a consequence of the last statement~\eqref{maximum_factor_statement_4} of Lemma~\ref{maximum_factor}, namely
\begin{align}\label{enter_condition}
(x_{\omega}^{\epsilon}(n-1),x_{\omega}^{\epsilon}(n))
\;\in\;
\mathfrak{R}_{\rm I}\times \mathfrak{R}_{\rm I}^c
\qquad\Longrightarrow\qquad x_{\omega}^{\epsilon}(n)\in\mathfrak{R}_{\rm II} 
\;.
\end{align}
Therefore, a rotation requires a stay in $\mathfrak{R}_{\rm II}$ at an integer-valued time and a later hit of $\mathfrak{R}_{\rm IV}$ at a half-integer-valued time, notably for all $N\in\mathbb{N}$ one has
\begin{align}\label{rotation_condition_1}
\begin{split}
&\exists\textnormal{ }0< M_1<M_2\leq N:\quad (x(0),x_{\omega}^{\epsilon}(M_1),x_{\omega}^{\epsilon}(M_2))
\;\in\;
\mathfrak{R}_{\rm I}\times \mathfrak{R}_{\rm I}^c\times \mathfrak{R}_{\rm I}
\\
&\qquad\qquad\qquad\qquad\qquad\Downarrow\\
& \exists\textnormal{ } 0< P\leq Q<N:\quad (x_{\omega}^{\epsilon}(P),x_{\omega}^{\epsilon}(Q+\tfrac{1}{2}))\in\mathfrak{R}_{\rm II} \times\mathfrak{R}_{\rm IV} 
\end{split}
\end{align}
with the understanding that $M_1$, $M_2$, $P$ and $Q$ are required to be integers. The passage through $\mathfrak{R}_{\rm III}$ in \eqref{rotation_condition_1} is first analyzed under the hypothesis of bounded support of $\kappa_\sigma$,~\textit{i.e.}, $C_4<\infty$. Lemma~\ref{no_large_forward_jump_bounded} states (under the latter assumption) that the dynamics can only leave $\mathfrak{R}_{\rm II}$ by entering a certain subset of $\mathfrak{R}_{\rm III}$. To formulate it precisely, we decompose~$\mathfrak{R}_{\rm III}$~into
\begin{align*}
&
\mathfrak{R}^<_{\rm III}
\,=\,\left\{x\in\mathbb{R}\;:\; \KConst \epsilon\leq x< 2(C_4)^2\KConst \epsilon \right\}
\;,
\\
&
\mathfrak{R}^>_{\rm III}
\,=\,\left\{x\in\mathbb{R}\;:\; 2(C_4)^2\KConst \epsilon\leq x\leq (\KConst {\epsilon})^{-1} \right\}
\;.
\end{align*}
%

\begin{lemma}
\label{no_large_forward_jump_bounded}
Assume $C_4<\infty$. Then for all $\epsilon>0$ sufficiently small and all $m\in\frac{1}{2}\mathbb{N}$ 
\begin{align}
\label{no_large_forward_jump_bounded_statement}
x(m)\in\mathfrak{R}_{\rm II}
\;\;\mbox{ and }\;\; x(\lceil m\rceil +\tfrac{1}{2})\not\in\mathfrak{R}_{\rm II} 
\qquad\Longrightarrow\qquad 
x(\lceil m\rceil+\tfrac{1}{2})\in\mathfrak{R}^<_{\rm III} \;,
\end{align}
where $\lceil m\rceil=\min\{n\in\NM\,:\,n\geq m\}$.
\end{lemma}

\noindent\textbf{Proof.} If $m\in\NM$, the statement is trivial because $C_4\geq \kappa$. Hence suppose $m\not\in\NM$ and set $x=x(m)$. Then $x(\lceil m\rceil +\tfrac{1}{2})=(DQ)\cdot x$. Now, since $x\in\mathfrak{R}_{\rm II}$, $Q\cdot x$ is positive and obeys
$$
Q\cdot x
\;=\;
\frac{(1+\epsilon^2\alpha)x+(a{-b}-\epsilon\beta)\epsilon}{1+\epsilon^2\delta-(a{+b}+\epsilon\gamma)\epsilon x}
\;<\;
\frac{\frac{3}{2}\,\KConst\,\epsilon}{\frac{3}{4}}\;=\;{2}\,\KConst\,\epsilon
\;.
$$
Thus, $(DQ)\cdot x = \kappa^2(Q\cdot x)\in (0,C_4^22\KConst\epsilon)$, and hence $(DQ)\cdot x\in\mathfrak{R}^<_{\rm III}$ due to $(DQ)\cdot x\not\in\mathfrak{R}_{\rm II}$.
\hfill $\square$

\subsection{The associated renewal process}
\label{sec-renewal}

Now let $\epsilon>0$ be sufficiently small so that Lemmata~\ref{maximum_factor}~and~\ref{no_large_forward_jump_bounded} hold. Lemma~\ref{no_large_forward_jump_bounded} implies a consequence of the lower statement of~\eqref{rotation_condition_1}, namely
\begin{align}\label{rotation_condition_bounded_1}
\begin{split}
& \exists\textnormal{ } 0< P\leq Q<N:\quad (x_{\omega}^{\epsilon}(P),x_{\omega}^{\epsilon}(Q+\tfrac{1}{2}))\in \mathfrak{R}_{\rm II}\times\mathfrak{R}_{\rm IV}\\
&\qquad\qquad\qquad\qquad\qquad\Downarrow\\
& \exists\textnormal{ } 1< R< S<N:\quad (x_{\omega}^{\epsilon}(R+\tfrac{1}{2}),x_{\omega}^{\epsilon}(S+\tfrac{1}{2}))\in \mathfrak{R}^<_{\rm III}\times\mathfrak{R}_{\rm IV}
\end{split}
\end{align}
for all $N\in\mathbb{N}$, with $P$, $Q$, $R$ and $S$ required to be integers. In view of statement~\eqref{maximum_factor_statement_2} of Lemma~\ref{maximum_factor}, the lower statement of~\eqref{rotation_condition_bounded_1} implies, in turn, 
\begin{align}\label{rotation_condition_bounded_2}
\begin{split}
& \exists\textnormal{ } 1< R< S<N:\quad (x_{\omega}^{\epsilon}(R+\tfrac{1}{2}),x_{\omega}^{\epsilon}(S+\tfrac{1}{2}))\in \mathfrak{R}^<_{\rm III}\times\mathfrak{R}_{\rm IV}\\
&\qquad\qquad\qquad\qquad\qquad\Downarrow\\
& \exists\textnormal{ } 2< N_1\leq N_2<N:\quad \prod\limits_{n=N_1}^{N_2}k\kappa_{\sigma_n}^2> \big(2C_4^2K^2\epsilon^2\big)^{-1}
\end{split}
\end{align}
for all $N\in\mathbb{N}$, where $R$, $S$, $N_1$ and $N_2$ are integers, since at least the width of $\mathfrak{R}^>_{\rm III}$ has to be overcome.

\vspace{.2cm}

As mentioned above, a rotation requires, in particular, the run of $x_{\omega}^{\epsilon}(\cdot)$ from $\mathfrak{R}_{\rm I}$ into $\mathfrak{R}_{\rm I}^c$ and then back to $\mathfrak{R}_{\rm I}$. If the starting and terminating region $\mathfrak{R}_{\rm I}$ were a singleton, the completion of a rotation would be construable as the occurence of a renewal of the process. Such renewals do not actually occur in the present process, since the locations of the dynamics after the re-entrances into $\mathfrak{R}_{\rm I}$ are vague. Accordingly, the random durations of the respective rotations are not identically distributed. However, the statements~\eqref{rotation_condition_1},~\eqref{rotation_condition_bounded_1},~\eqref{rotation_condition_bounded_2} combined imply
\begin{align}\label{rotation_condition_final_bounded}
\begin{split}
&\exists\textnormal{ }0< M_1<M_2\leq N:\quad (x(0),x_{\omega}^{\epsilon}(M_1),x_{\omega}^{\epsilon}(M_2))\in\mathfrak{R}_{\rm I}\times\mathfrak{R}_{\rm I}^c\times\mathfrak{R}_{\rm I}\\
&\qquad\qquad\qquad\qquad\qquad\Downarrow\\
& \exists\textnormal{ } 2< N_1\leq N_2<N:\quad \prod\limits_{n=N_1}^{N_2}k\kappa_{\sigma_n}^2> \big(2C_4^2K^2\epsilon^2\big)^{-1}\,.
\end{split}
\end{align}
Thus, these random durations can be uniformly bounded from below by i.i.d. (and $\mathbb{R}^+$-valued) random variables $\left\{X_n\right\}_{n\in\mathbb{N}}$ satisfying for all $s\geq 0$
\begin{align}\label{definition_interarrival_times_bounded}
\mathbf{P}(X_1\leq s)
\;=\;
\mathbf{P}\bigg(\exists\textnormal{ } 2< N_1\leq N_2< \lfloor s\rfloor:\prod\limits_{n=N_1}^{N_2}k\kappa_{\sigma_n}^2> \big(2C_4^2K^2\epsilon^2\big)^{-1}\bigg)
\;,
\end{align}
which are then proper \textit{interarrival times} and specify a \textit{renewal process} (see~\cite{GS}, Section~10)~via
\begin{align}\label{renewal_process}
\mathtt{P}(t)
\;=\;
\max\Big\{M\in\mathbb{N}: \sum_{n=1}^MX_n\leq t\Big\}\,,\qquad t\geq 0
\;.
\end{align}
Now, the interarrival times of the renewal process~\eqref{renewal_process} bound the random durations of the rotations from below. The renewal function $\langle \mathtt{P}(\cdot)\rangle$, accordingly, bounds the (expected) rotation number $\frac{1}{\pi}\left\langle\theta_{\omega}^{\epsilon}(N)\right\rangle$ from above. Indeed,~\eqref{rotation_condition_final_bounded},~\eqref{definition_interarrival_times_bounded}~and~\eqref{renewal_process} imply
\begin{align*}
\frac{1}{\pi}\left\langle\theta_{\omega}^{\epsilon}(N)\right\rangle
\;\leq\; 
\langle \mathtt{P}(N )\rangle\;+\;
\frac{1}{2}
\;,
\end{align*}
for any starting point $\theta_{\omega}^{\epsilon}(0)\in \left(0,\frac{\pi}{2}\right]$.
Hence the \textit{elementary renewal theorem} \cite[Section~10]{GS}
\begin{align}\label{renewal_theorem_statement}
\lim\limits_{t\rightarrow\infty}\frac{\langle \mathtt{P}(t)\rangle}{t} 
\;=\;
\langle X_1\rangle^{-1}
\end{align}
yields
\begin{align}\label{rotation_number_estimate}
\limsup\limits_{N\rightarrow\infty}\frac{1}{\pi}\frac{\langle\theta_{\omega}^{\epsilon}(N)\rangle}{N}
\;\leq \;
\langle X_1\rangle^{-1}
\;.
\end{align}
Thus the next aim is a lower bound of the mean of the interarrival time $X_1$.

\subsection{The large deviation estimate}
\label{sec-LD}

The present section is devoted to a lower bound on the expectation of $X_1$ given by \eqref{definition_interarrival_times_bounded}. The desired lower bound can be obtained by controlling the probability of the event
\begin{align}\label{event_in_large_deviation_estimate}
\exists\textnormal{ } 2< N_1\leq N_2<N:\quad \prod\limits_{n=N_1}^{N_2}k\kappa_{\sigma_n}^2> \big(2C_4^2K^2\epsilon^2\big)^{-1}
\end{align}
for $N\in\mathbb{N}$. A rough upper bound on the probability of~\eqref{event_in_large_deviation_estimate}, a union of events, is given by the sum of the probabilities of the single events, \textit{i.e.}, for fixed $N_1$ and $N_2$. This turns out to be sufficient for our purposes. As a preparation for bounding the probabilities of the single events, let us observe that there exists a unique {$\varrho_k\in (0,\frac{\nu}{2})$} such that
\begin{align}\label{existence_of_moment_1}
\left\langle \big(k\kappa^2_{\sigma}\big)^{\varrho_k}\right\rangle\;=\;1
\end{align}
(cf.~\cite{GGG}, Section 1.2) is satisfied. Indeed,~\eqref{existence_of_moment_1} is equivalent to $f(\varrho_k)=g_k(\varrho_k)$, where
$$
f:[0,\mbox{\small $\frac{\nu}{2}$}]\rightarrow (0,1]\,,\quad \varrho\mapsto \langle\kappa^{2\varrho}\rangle\,,\qquad\qquad\textnormal{and}\qquad\qquad g_k:[0,\mbox{\small $\frac{\nu}{2}$}]\rightarrow [k^{-\frac{\nu}{2}},1]\,,\quad\varrho\mapsto k^{-\varrho}\,.
$$
But $f$ is continuous, convex and obeys $f^{-1}(1)=\{0,\frac{\alpha}{2}\}$ and $g_k$ is bijective and decreasing. Moreover,~\eqref{existence_of_moment_1} implies that all $\xi\in(0,\varrho_k)$ satisfy 
$$
\left\langle \big(k\kappa^2_{\sigma}\big)^{\varrho_k-\xi}\right\rangle
\;<\;1
\;.
$$

\begin{lemma}\label{large_deviation_estimate}
For some $k>1$ let $\varrho_k\in (0,\frac{\nu}{2})$ be such that it satisfies~\eqref{existence_of_moment_1}. Then,
\begin{align}\label{large_deviation_estimate_statement}
\mathbf{P}\mbox{\small $\bigg(\exists\textnormal{ } 2< N_1\;\leq\; N_2<N:\prod\limits_{n=N_1}^{N_2}k\kappa^2_{\sigma_n}>\zeta^{-1}\bigg)$}\;\leq\; \zeta^{\varrho_k-\xi}N\Big(\left\langle\big(k\kappa_{\sigma_n}^2\big)^{\varrho_k-\xi} \right\rangle^{-1}-1\Big)^{-1}
\end{align}
holds for all $N\in\mathbb{N}$, $\xi\in(0,\varrho_k)$ and $\zeta>0$.
\end{lemma}

\noindent\textbf{Proof.} The series of estimates
\begin{align*}
\begin{split}
\mathbf{P}\mbox{\small $\left(\exists\textnormal{ }2<N_1\;\leq\; N_2<N: \textnormal{ }\prod\limits_{n=N_1}^{N_2}k\kappa_{\sigma_n}^2> \zeta^{-1}\right)$}
&\;\leq\; \sum\limits_{N_2=3}^{N-1}\mathbf{P}\mbox{\small $\left(\exists\textnormal{ }2<N_1 \leq N_2:\textnormal{ } \prod\limits_{n=N_1}^{N_2}k\kappa_{\sigma_n}^2>\zeta^{-1}\right)$}\\
&\;\leq\; \sum\limits_{N_2=3}^{N-1}\sum\limits_{N_1=3}^{N_2}\mathbf{P}\mbox{\small $\left(\prod\limits_{n=N_1}^{N_2}k\kappa_{\sigma_n}^2>\zeta^{-1}\right)$}\\
&\;\leq\; \sum\limits_{N_2=3}^{N-1}\sum\limits_{N_1=3}^{N_2}\mathbf{P}\mbox{\small $\left(\prod\limits_{n=N_1}^{N_2}\big(k\kappa_{\sigma_n}^2\big)^{\varrho_k-\xi}>\zeta^{\xi-\varrho_k}\right)$}\\
&\;\leq\; \zeta^{\varrho_k-\xi}\sum\limits_{N_1=3}^{N-1}\sum\limits_{N_1=3}^{N_2}\left\langle\big(k\kappa_{\sigma_n}^2\big)^{\varrho_k-\xi} \right\rangle^{N_2-N_1+1}\\
&\;\leq\; \zeta^{\varrho_k-\xi}N\sum\limits_{n\in\mathbb{N}}\left\langle\big(k\kappa_{\sigma_n}^2\big)^{\varrho_k-\xi} \right\rangle^{n}\\
&\;=\;\zeta^{\varrho_k-\xi}N\Big(\left\langle\big(k\kappa_{\sigma_n}^2\big)^{\varrho_k-\xi} \right\rangle^{-1}-1\Big)^{-1}
\end{split}
\end{align*}
completes the proof.
\hfill $\square$

\vspace{0.2cm}

The desired lower bound on $\langle X_1\rangle$ is now obtained by using the estimate proved in Lemma~\ref{large_deviation_estimate}.

\begin{lemma}\label{lower_bound_expectation_interarrival_time_bounded}
{For some $k>1$ let $\varrho_k\in (0,\frac{\nu}{2})$ be such that it satisfies~\eqref{existence_of_moment_1}. Moreover, let $\xi\in(0,\varrho_k)$.}
Then, sufficiently small $\epsilon>0$ satisfy the estimate
\begin{align}\label{lower_bound_expectation_interarrival_time_bounded_statement}
\langle X_1\rangle
\;\geq\; 
\frac{1}{2}(2C_4^2K^2\epsilon^2)^{\xi-\varrho_k}\left(1-\left\langle\big(k\kappa_{\sigma_n}^2\big)^{\varrho_k-\xi} \right\rangle\right)\,.
\end{align}
\end{lemma}

\noindent\textbf{Proof.}  Let $N\in\mathbb{N}$.
In view of~\eqref{definition_interarrival_times_bounded} and the bound~\eqref{large_deviation_estimate_statement} obtained in Lemma~\eqref{large_deviation_estimate}, it holds that
$$
\mathbf{P}(X_1\leq N)
\;\leq\; (2C_4^2K^2\epsilon^2)^{\varrho_k-\xi}N\Big(\left\langle\big(k\kappa_{\sigma_n}^2\big)^{\varrho_k-\xi} \right\rangle^{-1}-1\Big)^{-1}\,.
$$
Thus, setting
$$
\Upsilon\;=\;(2C_4^2K^2\epsilon^2)^{\xi-\nu_c}\Big(\left\langle\big(k\kappa_{\sigma_n}^2\big)^{\varrho_k-\xi} \right\rangle^{-1}-1\Big)\,,
$$
it holds that
\begin{align*}
\begin{split}
\langle X_1\rangle&\;=\;\sum\limits_{N\in\mathbb{N}}\mathbf{P}\left(X_1\geq N\right)\;=\;\sum\limits_{N\in\mathbb{N}_0}\left[1-\mathbf{P}\left(X_1\leq N\right)\right]\;\geq \;\sum\limits_{N\in\mathbb{N}_0}\min\left\{0,1-\Upsilon^{-1} N\right\}\\
&\;=\;\sum\limits_{N=0}^{\left\lfloor\Upsilon\right\rfloor}\left[1-\Upsilon^{-1} N\right]\;=\;\left\lfloor\Upsilon\right\rfloor-\Upsilon^{-1}\sum\limits_{N=1}^{\left\lfloor\Upsilon\right\rfloor}N\;=\;\left\lfloor\Upsilon\right\rfloor-\frac{\lfloor\Upsilon\rfloor\left(\lfloor\Upsilon\rfloor+1\right)}{2\Upsilon}\;\geq\; \frac{(\Upsilon-1)^2}{2\Upsilon}\,,
\end{split}
\end{align*}
which implies~\eqref{lower_bound_expectation_interarrival_time_bounded_statement} for sufficiently small $\epsilon$.
\hfill $\square$

\subsection{Conclusion of the argument: the case of bounded support}
\label{sec-bounded}

Now all technical elements needed for the proof of Theorem~\ref{main_result} are prepared. The following result also includes the generalization to arbitrary hyperbolic critical energies.

\begin{theo}\label{upper_bound_rotation_number_bounded}
Let $E_c$ be a hyperbolic critical energy of a random polymer Hamiltonian. Let $C_1$ to $C_4$ be finite positive constants. Suppose that $\gamma^0<0$ and that the exponent $\nu>0$ is defined by \eqref{eq-KappaCrit}, namely $\langle \kappa_\sigma^{\nu}\rangle=1$. For all $\delta\in (0,\nu)$ there exist $C_\delta<\infty$ such that sufficiently small $\epsilon$ satisfy
$$
|\Nn(E_c+\epsilon)\,-\,\Nn(E_c)|
\;\leq\;
C_\delta\hspace{0.5mm}\epsilon^{\nu-\delta}
\;.
$$
\end{theo}

\noindent\textbf{Proof.}
Due to \eqref{eq-IDSRotnumber} it is sufficient to prove a bound on the rotation number.
For $k>1$ let $\varrho_k\in (0,\frac{\nu}{2})$ be such that it satisfies~\eqref{existence_of_moment_1} and $\xi\in (0,\varrho_k)$. Furthermore, let $\epsilon>0$ be such that the statements of Lemma~\ref{maximum_factor}, Lemma~\ref{no_large_forward_jump_bounded} and Lemma~\ref{lower_bound_expectation_interarrival_time_bounded} hold. Then,~\eqref{rotation_number_estimate}~and~\eqref{lower_bound_expectation_interarrival_time_bounded_statement} imply
\begin{align}\label{upper_bound_rotation_number_bounded_1}
\limsup\limits_{N\rightarrow\infty}\frac{1}{2\pi}\frac{\langle\theta_{\omega}^{\epsilon}(N)\rangle}{N}\;\leq\; 2\hspace{0.5mm}(2C_4^2K^2\epsilon^2)^{\varrho_k-\xi}\left(1-\left\langle\big(k\kappa_{\sigma_n}^2\big)^{\varrho_k-\xi} \right\rangle\right)^{-1}\,.
\end{align}
But $\varrho_k$ is continuous in $k$ and converges to $\frac{\nu}{2}$ as $k\downarrow 1$. Thus the r.h.s. of \eqref{upper_bound_rotation_number_bounded_1} is bounded above by $C_\delta\epsilon^{\nu-\delta}$ for
$$
C_\delta\;=\;2\;(2C_4^2K^2)^{\varrho_k-\xi}\left(1-\left\langle\big(k\kappa_{\sigma_n}^2\big)^{\varrho_k-\xi} \right\rangle\right)^{-1}\,,
$$
where $k>1$ and $\xi\in (0,\varrho_k)$ have to be chosen such that $2(\varrho_k-\xi)=\nu-\delta$ holds.
\hfill $\square$

\subsection{The case of unbounded support}
\label{sec-unbounded}

Proving upper bounds on the rotation number is somewhat more involved, once the assumption~$C_4<\infty$ is dropped. In this situation, there does not exists some $\mathtt{K}\in (1,\infty)$ such that
\begin{align}\label{good_event}
\kappa_{\sigma_1}\;\leq\; \mathtt{K}\,,\quad \kappa_{\sigma_2}\;\leq\; \mathtt{K}\,,\quad\dots\,,\quad\kappa_{\sigma_N}\;\leq\;  \mathtt{K}
\end{align}
holds with probability $1$. However, the above arguments can be applied to the cases where the event~\eqref{good_event} does occur. Thus,~\eqref{rotation_condition_final_bounded} reads more generally 
\begin{align}\label{rotation_condition_final_unbounded}
\begin{split}
&\exists\textnormal{ }0< M_1<M_2\leq N:\quad (x(0),x_{\omega}^{\epsilon}(M_1),x_{\omega}^{\epsilon}(M_2))\in\mathfrak{R}_{\rm I}\times\mathfrak{R}^{c}_{\rm I}\times\mathfrak{R}_{\rm I}\\
&\qquad\qquad\qquad\qquad\qquad\Downarrow\\
& \eqref{good_event}\textnormal{ is violated}\quad\mbox{or}\quad\exists\textnormal{ } 2< N_1\leq N_2<N:\quad \prod\limits_{n=N_1}^{N_2}k\kappa_{\sigma_n}^2> \big(2\mathtt{K}^2K^2\epsilon^2\big)^{-1}\,.
\end{split}
\end{align}
Thus, let us analyze the renewal process induced by the i.i.d. interarrival times $\{\widetilde{X_n}\}_{n\in\mathbb{N}}$ with
\begin{align*}
\begin{split}
\mathbf{P}\big(\widetilde{X_1}\leq s\big)\;=\;\mathbf{P}\bigg(\eqref{good_event}\textnormal{ is violated}\quad\mbox{or}\quad\exists\textnormal{ } 2< N_1\leq N_2<N:\quad \prod\limits_{n=N_1}^{N_2}k\kappa_{\sigma_n}^2> \big(2\mathtt{K}^2K^2\epsilon^2\big)^{-1}\bigg)
\;,
\end{split}
\end{align*}
where $s\geq 0$ and $N=\lfloor s\rfloor$, instead of~\eqref{renewal_process}. Now, the probability of the violation of~\eqref{good_event} is dealt with by
\begin{align*}
\mathbf{P}\left(\hspace{0.5mm}\eqref{good_event}\textnormal{ is violated}\right)\leq N\;\mathbf{P}(\kappa >\mathtt{K})\;=\;N\;\mathbf{P}(\kappa^{\nu} >\mathtt{K}^{\nu})\;\leq\; \mathtt{K}^{-\nu}N\,,
\end{align*}
where we used $\langle\kappa^{\nu}_{\sigma}\rangle=1$, so that Lemma~\ref{large_deviation_estimate} implies that all $k>1$ and $\xi\in (0,\varrho_k)$~obey 
\begin{align}\label{upper_bound_probability_modified}
\mathbf{P}\big(\widetilde{X_1}\leq N\big)\;\leq\; N\left[\mathtt{K}^{-\nu}+(2\mathtt{K}^2K^2\epsilon^2)^{\varrho_k-\xi}\Big(\left\langle\big(k\kappa_{\sigma_n}^2\big)^{\varrho_k-\xi} \right\rangle^{-1}-1\Big)^{-1}\right]\,.
\end{align}
Clearly, the choice $\mathtt{K}=\epsilon^{-\frac{1}{2}}$ optimizes the order of the right side of~\eqref{upper_bound_probability_modified}  in $\epsilon$ as $k\downarrow 1$. This allows to prove that the bound in Theorem~\ref{upper_bound_rotation_number_bounded}  remains valid if the exponent $\nu-\delta$ is replaced by $\frac{\nu}{2}-\delta$, even if $C_4$ is not finite.

\vspace{.3cm}

\noindent {\bf Acknowledgements:}  We thank G\"unter Stolz for bringing the random dimer hopping model of Section~\ref{sec-intro} and its connection to spin chains to our attention. F.~D. received funding from the {\it Studienstiftung des deutschen Volkes}. This work was also partly supported by the~DFG.


\end{document}